\documentclass{PoS}

\title{ Array configuration studies for the Square Kilometre Array
-- Implementation of figures of merit based on spatial dynamic range}

\ShortTitle{SKADS: Array Configuration Studies}

\author{Dharam Vir Lal
         \thanks{This work has been supported by the
European Community Framework Programme 6,
Square Kilometer Array Design Studies (SKADS), contract no 011938.}\\
        Max-Planck-Institut f\"ur Radioastronomie,
        Auf dem H\"ugel 69, 53121 Bonn, Germany\\
        E-mail: \email{dharam@mpifr-bonn.mpg.de}}

\author{Andrei P. Lobanov\\
        Max-Planck-Institut f\"ur Radioastronomie,
        Auf dem H\"ugel 69, 53121 Bonn, Germany\\
        E-mail: \email{alobanov@mpifr-bonn.mpg.de}}

\author{Sergio Jim\'enez-Monferrer\\
Max-Planck-Institut f\"ur Radioastronomie, 
        Auf dem H\"ugel 69, 53121 Bonn, Germany\\
and \\
Universidad de Valencia,
Dr. Moliner 50, 46100 Burjassot, Valencia, Spain\\
E-mail: \email{sergio.jimenez@uv.es}}

\abstract{The Square Kilometre Array (SKA) will be operating at the
time when several new large optical, X-ray and Gamma-ray facilities
are expected to be working.  To make SKA both competitive and
complementary to these large facilities, thorough design studies are
needed, focused in particular on imaging performance of the array. One
of the crucial aspects of such studies is the choice of the array
configuration, which affects substantially the resolution, rms noise,
sidelobe level and dynamic range achievable with the SKA.  We present
here a quantitative assessment of the effect of the array
configuration on imaging performance of the SKA, introducing the
spatial dynamic range (SDR) and a measure of incompleteness of the Fourier
domain coverage ($\Delta u/u$) as prime figures of merit.}

\FullConference{}

\begin{document}

\section{Introduction}
\label{intro}

The demanding breadth of the science case and technical specifications
of the Square Kilometre Array (SKA) makes the design of the array a
complex, multi-dimensional undertaking (Jones 2003).  Although a
number of factors may limit the performance of the SKA, including
radio frequency interference, atmospheric and ionospheric effects
(Perley 1999, Thompson et al. 1986), the array configuration is one of
the most critical aspects of the instrument design (ISPO 2006), as it
will be very difficult to modify or modernise the station locations
after the construction phase has finished.

Several approaches have been exercised in order to obtain an
optimised array configuration (Conway 1998, Bregman 2000,
Conway 2000a, 2000b, Kogan 2000a, 2000b, Noordam 2001, Cohanim et al. 2004,
Bregman 2005, Kogan \& Cohen 2005, Lonsdale 2005, Morita \& Holdaway 2005).
Judgement of quality and fidelity of synthesised images is typically made by
estimating the "dynamic range", the ratio between the peak brightness
on the image and the r.m.s. noise in a region free of emission of the image
(Perley 1999).  High dynamic range is necessary for
imaging a high-contrast feature, which is a key requirement for the
SKA.  SKA design goal for the synthesised image dynamic range is
10$^6$ at 1.4 GHz (Wright 2002, 2004).

Alongside the dynamic range, one of the basic figures of merit (FoM)
characterising imaging performance of an interferometer is the spatial
dynamic range (SDR) quantifying the range of spatial scales that can
be reconstructed from interferometer data (Lobanov 2003).  The SDR of
an interferometer depends on a number of instrument parameters,
including the integration time of the correlator, channel bandwidth,
and the coverage of the Fourier domain ({\em uv}-coverage).  The
latter factor poses most stringent constraints on the design,
particularly for arrays with a relatively small number of elements.
The quality of the {\em uv}-coverage can be expressed by the $uv$-gap
parameter or $\Delta u/u$, characterising the relative size of "holes"
in the Fourier plane.  Basic analytical estimates indicate that the
SKA should have $\Delta u/u$ $\lesssim$ 0.2 (Lobanov 2003) over the
entire range of baselines to provide sufficient imaging capabilities
and warrant that the SDR of the SKA would not be {\em uv}-coverage
limited.

The SKA configuration must provide {\em uv}-coverages that satisfy
several key requirements derived from the prime science goals of the
instrument: (i)~good snapshot and deep imaging over 1 degree field of
view, (ii)~high brightness temperature (T$_{\rm b}$) for extended
objects, (iii)~dense core for transients$/$pulsar$/$SETI, and
(iv)~long baselines for milli-arcsecond imaging.  The combination of
these requirements with the benchmark figures for the dynamic range of
continuum (10$^6$) and spectral line (10$^5$) observations poses a
substantial challenge for the array design and for the antenna
distribution in particular.

A commonly used approach to designing the antenna configuration for an
interferometric array relies on optimising the {\em uv}-coverage
by minimising sidelobes or providing a desired beam shape (cf.,
Cornwell 1986, Kogan 2000a, 2000b). This approach assumes implicitly that the
field of view is not crowded and the target objects are marginally
resolved (so that the structural information can be recovered
efficiently even if a substantial fraction of spatial frequencies is
undersampled). Neither of these two assumptions will be correct for
the SKA operating in the $\mu$Jy regime. This complication
requires additional constraints and considerations to be employed in
order to warrant successful imaging of all spatial scales sampled by
the interferometer.

This issue can be addressed effectively by ensuring a constant $\Delta
u / u$ over the entire range of baselines, which will provide equal
sensitivity to all spatial scales sampled by an interferometer and 
realize the full spatial dynamic range of the instrument.

\section{Spatial dynamic range of an interferometer}

Spatial dynamic range of an interferometer is determined by several
major factors related to details of signal processing, design of the
primary receiving element, and distribution of the collecting area of
an array (Lobanov 2003).  The maximum achievable spatial dynamic range
(${\rm SDR_{\rm FoV}}$), is given by the ratio of the field of view and
the synthesised beam (HPBW).  For an array composed of parabolic
antennas
$$
{\rm SDR_{\rm FoV}} \approx 0.80 \frac{{\rm B_{max}}}{\eta_{\rm a}^{0.5} {\rm D}},
$$
where, ${\rm B_{max}}$ is the longest baseline in the array,
$\eta_{\rm a}$ is the aperture efficiency, and D is the diameter of
the antenna.  However, ${\rm SDR_{\rm FoV}}$ can be typically achieved
only at the shortest baselines.  The SDR is significantly reduced at
the highest instrumental resolution, due to finite bandwidth and
integration time, and incomplete sampling of the Fourier
domain.  Applying an averaging time of $\tau_{\rm a}$ to
interferometric data limits the maximum size $\theta_{\rm a}$, of
structure detected at full sensitivity to $\theta_{\rm a} =
c\,(\nu_{\rm obs} \omega_{\rm e} \tau_{\rm a} B_{\rm max})^{-1}$
(Bridle \& Schwab 1999), where $\nu_{\rm obs}$ is the observing frequency and
$\omega_{\rm e}$ is the angular rotation speed of the Earth. Relating
this size to the HPBW of the interferometer yields the maximum SDR
that can be achieved at a given integration time:
$$
{\rm SDR_\tau} \approx (\omega_{\rm e}\, \tau_{\rm a})^{-1} \approx 1.13 \times 10^4 \tau^{-1}.
$$
Bandwidth smearing due to a finite fractional bandwidth, $\Delta\nu$,
leads to a reduction in the peak response $R_{\Delta\nu} = (1+ \Delta\nu\, 
\theta_\nu/\theta_{\rm HPBW})^{-1/2}$ (assuming a Gaussian bandpass and circular Gaussian tapering; see Bridle \& Schwab 1999). The resulting limit on spatial dynamic range, ${\rm SDR}_{\Delta\nu}$, can be approximated by the ratio $\theta_\nu/\theta_{\rm HPBW}$, which gives
$$
{\rm SDR_{\Delta\nu}} \approx {\Delta\nu}^{-1} (R_{\Delta\nu}^{-2} - 1)^{1/2},
$$
with $R_{\Delta\nu} \le 0.75$ typically assumed.  Finally, incomplete
sampling of the Fourier space yields ${\rm SDR}_{\Delta u} < {\rm
SDR}_{\rm FOV}$. The magnitude of the SDR reduction can be expressed
in terms of the "{\em uv}-gap" parameter, $\Delta u /u$, that can be
defined as follows: $\Delta u/u = (u_2 - u_1)/u_1$, where $u_1$, $u_2$
($u_2>u_1$) are the $uv$-radii of two adjacent baselines (sampling
respective structural scales $\theta_{1,2} =1/u{1,2}$, with $\theta_1 >
\theta_2$).  Applying the assumptions used for deriving ${\rm
SDR}_{\Delta\nu}$ gives a synthesised beam that can be
well-approximated by a two-dimensional Gaussian. Then, for structures
partially resolved at $u_2$, the smallest resolvable size (or
variation of the size) can be estimated by requiring that a difference
in visibility amplitudes $V(u_1)$ and $V(u_2)$ can be detected at a given
SNR. This approach is similar to the one applied to
determining resolution limits of an interferometer ({\em c.f.},
Lobanov et al. 2001, Lobanov 2005), and for $\mathrm{SNR}\gg 1$ it yields
$\theta_2/\theta_1 = (\pi/4) [\ln 2\, \ln({\rm
SNR})]^{-1/2}$. The ratio
$\theta_{2}/\theta_1$ can be represented by the term $1 +
\Delta u/u$. With this term, the expression $ {\rm SNR}_{\Delta u} =
\exp [\pi^2 (1 + \Delta u /u)^2 / (16\,\ln 2)]$ gives the relation
between $\Delta u/u$ and the SNR required for detecting emission on
spatial scales corresponding to the ($u_1,\, u_2$) range. In case of a
filled aperture, for which $\Delta u/u \rightarrow 0$, the
corresponding ${\rm SNR}_{\Delta u}= {\rm SNR}_{\Delta u/u =0} = \exp[\pi^2/(16\,\ln
2)]$ and ${\rm SDR}_{\Delta u} = {\rm SDR}_{\rm FOV}$. For partially
filled apertures, the ratio ${\rm SDR}_{\Delta u}/{\rm SDR}_{\rm FOV}$
can be estimated from the ratio ${\rm SNR}_0/{\rm SNR}_{\Delta u}$,
which then gives
\[
{\rm SDR}_{\Delta u} = {\rm SDR}_{\rm FoV}/ \exp \left[ \frac{\pi^2}{16~{\rm ln}~2}\frac{\Delta u}{u}\left(\frac{\Delta u}{u}+2 \right) \right]\, .
\]
Strictly speaking, the $uv$-gap parameter is a function of the
location $(u,\,\theta)$ in the $uv$-plane (with $\theta$ describing
the position angle), and it should be represented by a density field
in the $uv$-plane. We will apply this description for making
assessments of realistic {\rm uv}-coverages obtained from our
simulations.  

The $1/\exp$ factor in the expression for $\mathrm{SDRuv}$,
calculated for two {\em uv}-points $u_1$ and $u_2$, essentially
provides an estimate of a fraction of power that can be recovered
between the respective angular scales ($\theta_2, \theta_1$) from the
sky brightness distribution. When an average value of $\Delta u/u$ over the
entire {\em uv}-coverage is determined, the $1/\exp$ factor can be taken
as a measure of ratio between the largest detectable structure and the
primary beam (FOV) of individual array elements (under condition that
the largest detectable size obtained from $\Delta u/u$ is smaller that the 
largest angular scale given by $1/u_\mathrm{min}$).

For an idealised, circular {\em uv}-coverage obtained
with a regular array ({\it i.e.} logarithmic-spiral) with $N$ stations
organised in $M$ arms extending over a range of baselines $(B_{\rm
min}, B_{\rm max})$, the $uv$-gap can be approximated by $\Delta u/u
\approx (B_{\rm max}/B_{\rm min})^{\xi}-1$, with $\xi = M/N$ for
baselines between antennas on a single arm, and $\xi = 1/N$ for all
baselines.  For instruments with multi-frequency synthesis (MFS)
implemented, $\Delta u/u$ should be substituted by $\Delta u/u -
\Delta\nu_{\rm mfs}$, where, $\Delta\nu_{\rm mfs}$ is the fractional
bandwidth over which the multi-frequency synthesis is being performed.
It should be noted that MFS will be not as effective improving $\Delta u/u$ 
for sources near the equator.

In real observations, the actual SDR is determined by the most
conservative of the estimates provided above.  The simulations
presented below will focus on ${\rm SDR}_{\Delta u}$, assuming
implicitly that the instrument is designed so that ${\rm SDR}_{\Delta
u} \ge {\rm SDR_{\Delta\nu}}$ and ${\rm SDR}_{\Delta u} \ge {\rm
SDR}_{\tau}$. Under this assumption, we investigate the relation
between different array configurations as described by $\Delta u/u$
and resulting image properties described by the rms noise, dynamic range,
and structural sensitivity of the array.

\section{Relation between {\em uv}-coverage and imaging capabilities of an interferometer}

\noindent
In order to provide a quantitative measure of the effect of array
configuration on imaging performance of an interferometer, we simulate
a set of an idealised array configurations, each providing
$$
\frac{\Delta u}{u} \equiv {\rm const}
$$
over the entire range of $u$ and $\theta$. 
 It should be
noted that these configurations are introduced
solely for the purpose of analysing the dependence of SDR on the
{\em uv}-coverage, and they are not intended to serve as
prototype configurations for the SKA 

We generate the test array
configurations by considering an equiangular, planar logarithmic spiral and
projecting this spiral on Earth's surface and determine the locations of
individual stations ({\it i.e.}, latitudes and longitudes) using World
GEOD system 1984 (Heiskanen \& Moritz 1967). We apply Halley's third-order
formula (Fukushima 2006), a modification of Borkowski's method
(Borkowski 1989), to transform Cartesian to geodetic coordinates.
This method is sufficiently precise; the maximum error of the latitude
and the relative height is less than 6~$\mu$arcsec for the range of
height, $-$10~km $\le$ height $\le$ 30,000 km, and is stable in the
sense that it converges for all coordinates including the
near-geocentre region and near-polar axis region.  We then transform
these coordinates to obtain ECEF (Earth Centered Earth Fixed)
coordinates using a glish script, and produce an input array
configuration file. The actual dimensions of the simulated array are
chosen such that baseline non-coplanarity is negligible and no
$w$-term correction is required. This does not affect conclusions
obtained from the simulations, which are generic and can be applied to
evaluate imaging performance of any interferometric array of
arbitrary configuration and extent.

\begin{table}
\caption{Telescope settings for the generation of a
range of visibility datasets.}
\begin{center}
\begin{tabular}{l|l}
Parameter         &   Value             \\ 
\hline
Frequency         &   L band (1.4 GHz)            \\ 
Antenna           &   SEFD$=335$\,K$^\dag$        \\ 
Bandwidth         &   3.2 MHz                     \\ 
No. of channels   &   1                  \\ 
Direction (J2000) &   00:00:00 $+$90.00.00 \\ 
Elevation limit   &   12 deg                \\ 
Shadow limit      &   0.001$^\ddag$        \\ 
Start\_time (IAT) &   22/08/2007 / 06:00   \\ 
Stop\_time (IAT)  &   22/08/2007 / 18:00
\end{tabular}
\end{center}
Note: $\dag$ -- similar to the system independent flux density (SEFD) of a
VLA antenna; $\ddag$ -- shadow limit is set such as not to constrain the
{\em uv}-coverages obtained.
\label{parameter1}
\end{table}

\begin{table}
\caption{Input group of source components used for the generation of simulated visibility datasets.}
\begin{center}
\begin{tabular}{c|cc|r}
   Source size & RA & Dec & Flux density \\
               &\multicolumn{2}{c|}{(J2000)} & \multicolumn{1}{c}{(Jy)} \\
\hline
 $0.1^{\prime} \times 0.1^{\prime}$ &03:00:00 &88.00.00 & 8.0 \\
 $1.0^{\prime} \times 1.0^{\prime}$ &08:00:00 &88.30.00 & 4.0 \\
 $1.2^{\prime} \times 0.4^{\prime}$ &16:00:00 &89.00.00 & 1.2 \\
 $3.0^{\prime} \times 1.0^{\prime}$ &21:00:00 &88.00.00 & 3.0 \\
 $12.0^{\prime} \times 4.0^{\prime}$ &06:00:00 &88.00.00 & 12.0 \\
 $42.0^{\prime} \times 14.0^{\prime}$ &18:00:00 &88.00.00 & 42.0 \\
 $120.0^{\prime} \times 40.0^{\prime}$ &00:00:00 &90.00.00 & 120.0 \\
\end{tabular}
\end{center}
\label{parameter2}
\end{table}

In order to generate {\em uv}-coverages satisfying the condition
$$
\frac{\Delta u}{u} (u,\,\phi) \equiv {\rm const},
$$
we place the array center at the North Pole, and consider a fiducial
field centered at a 90~deg declination (which yields circular {\em
uv}-coverages that can be described by a single value of $\Delta
u/u$).  The corresponding visibility datasets have been generated in
aips$++$ (ver 1.9, build 1556; {\tt casa.nrao.edu}).  Imaging has been
done with {\sc aips} package ({\tt www.aips.nrao.edu}).  Several
Fortran programs and glish scripts have been developed and used to
automate the process and provide a structured, uniform, repeatable and
robust processing.

We generate datasets for full
track observations (IAT 6:00--18:00 hrs), using a range of integration
times at a frequency of 1.4~GHz and with a 3.2~MHz bandwidth (see
Table~\ref{parameter1}).  Antenna sensitivity (listed in
Table~\ref{parameter1}) are assumed to be similar to a dish of Very
Large Array (VLA). The simulator does not take into account the effect
of the primary beam on imaging.  An elevation limit (horizon mask) of
12~deg and a shadow limit (maximum fraction of geometrically shadowed
area before flagging occurs) of 0.1 per~cent was introduced.

We produce a number of array configurations corresponding to different
values of $\Delta u/u$ and use them to obtain visibility datasets for
the chosen model brightness distribution (see Table~\ref{parameter2}).
We repeat this exercise for all generated array configurations and
perform identical pipeline analysis, to ensure a self-consistent
comparison of basic characteristics of the resulting dirty and
CLEAN-ed images.

\subsection{Simulations}

This section describes simulations and analysis of array configurations providing 
equal {\em uv}-gap at all baseline lengths.

\begin{figure}
\begin{center}
\includegraphics[height=6.5cm]{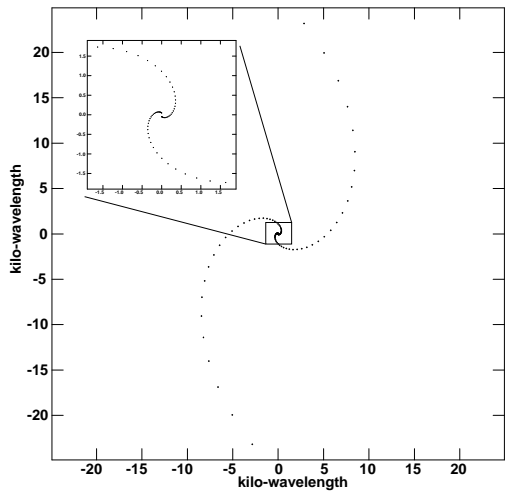} \\
\end{center}
\caption{ An example of a simulated 2-minute long snapshot {\em
uv}-coverage outlining relative locations of antenna stations.  The
largest baseline length, B$_{\rm max}= 5$~km, and the number of
antennas is $N=50$; In the simulations, only baselines to the central
station are considered, resulting in $\Delta u /u \equiv 0.19$ for a
full track observation.}
\label{uv45}
\end{figure}

We have tried two different algorithms for generating array
configurations with a given $\Delta u/u = const$ for all
baselines. Both algorithms have employed a logarithmic spiral
geometry, but differed in the realisation of changing the
characteristic value of $\Delta u/u$ from one configuration to
another.  In the first alogrithm, the total number of stations was
kept constant, while the baseline spread was gradually increased. In
the second approach, the baseline spread was kept constant, and
changing $\Delta u/u$ was achieved by changing the total number of
antennas in the array. We have found that the second approach is
superior for maintaining a constant noise level for different array
configurations, and we adopted it as the basis for our simulations.
In this set of simulations, the largest baseline length, $B_{\rm max}
= 5$~km is kept constant and different values of $\Delta u/u$ are
realized by varying the total number of antennas $N$ located on a
single arm of an equiangular, logarithmic spiral.  Fig.~\ref{uv45}
gives an example of an simulated snapshot {\em uv}-coverage generated
using this approach.

\begin{figure}
\begin{center}
\begin{tabular}{lll}
\includegraphics[height=14.2cm]{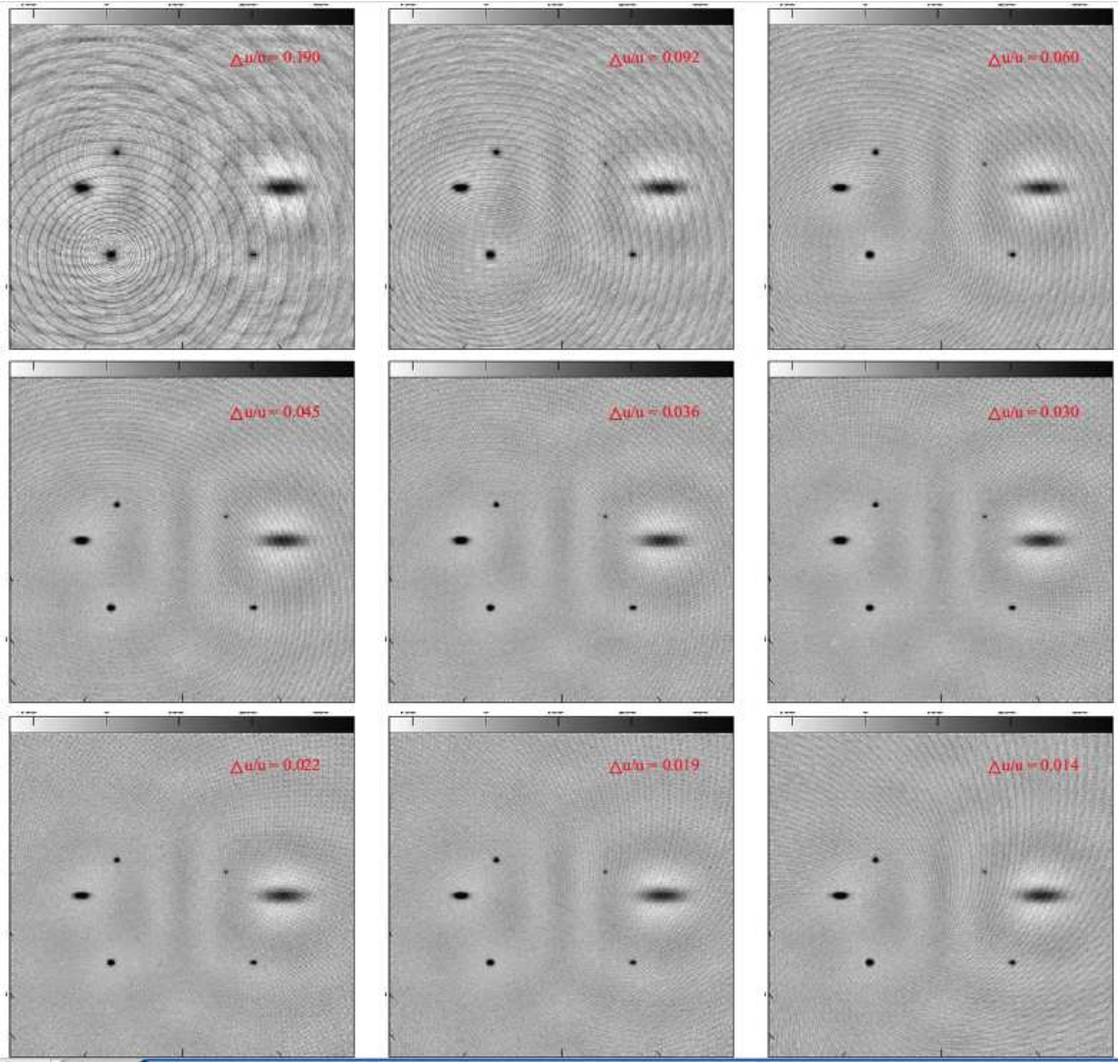} \\ [-0.6cm]
\end{tabular}
\end{center}
\caption{Dirty images obtained from the simulated data. 
Each panel is marked with the value of $\Delta u/u$ 
for the respective simulated dataset.}
\label{map12}
\end{figure}

\begin{figure}
\begin{center}
\begin{tabular}{lll}
\includegraphics[height=4.7cm]{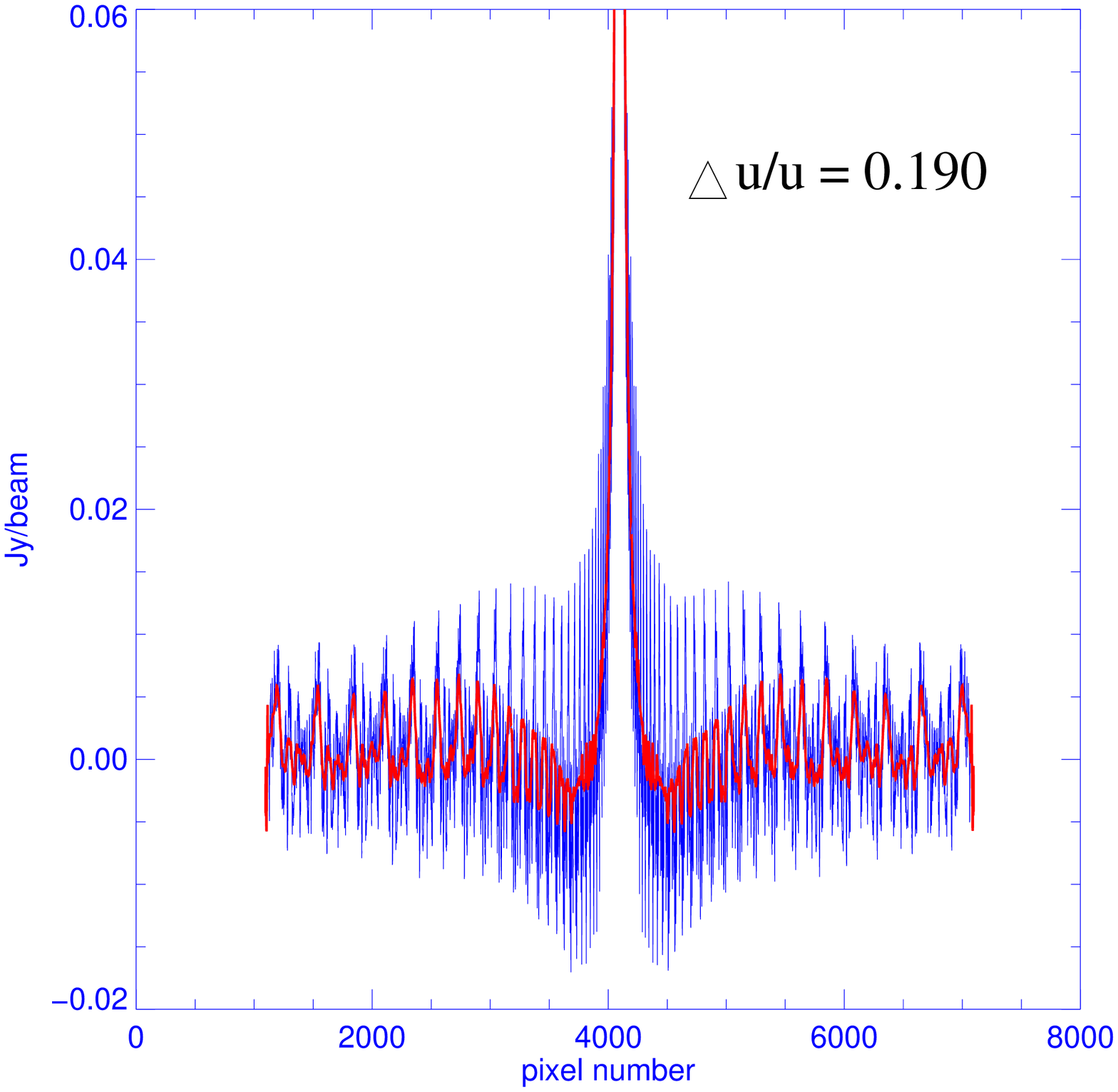} &
\includegraphics[height=4.7cm]{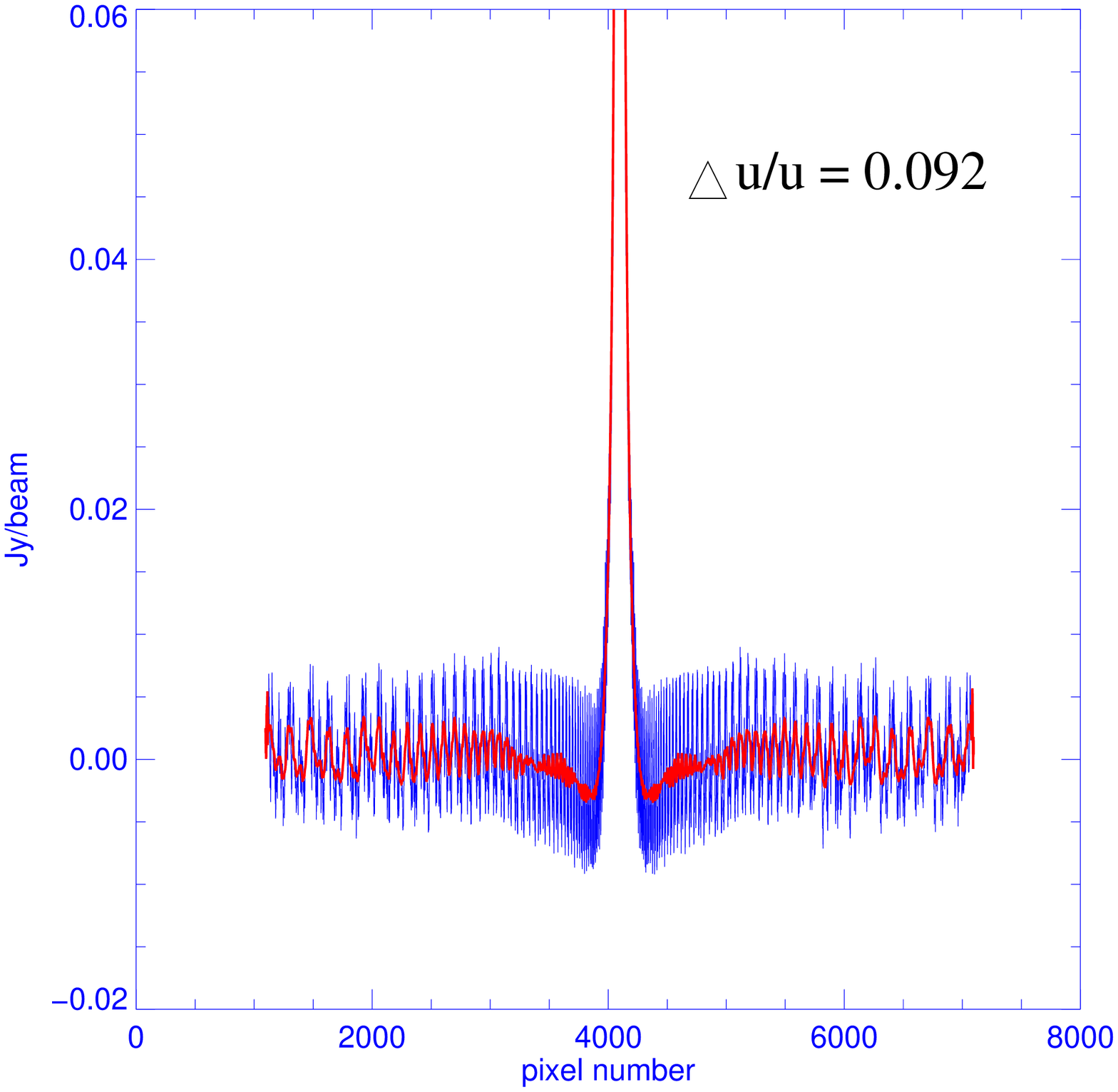} &
\includegraphics[height=4.7cm]{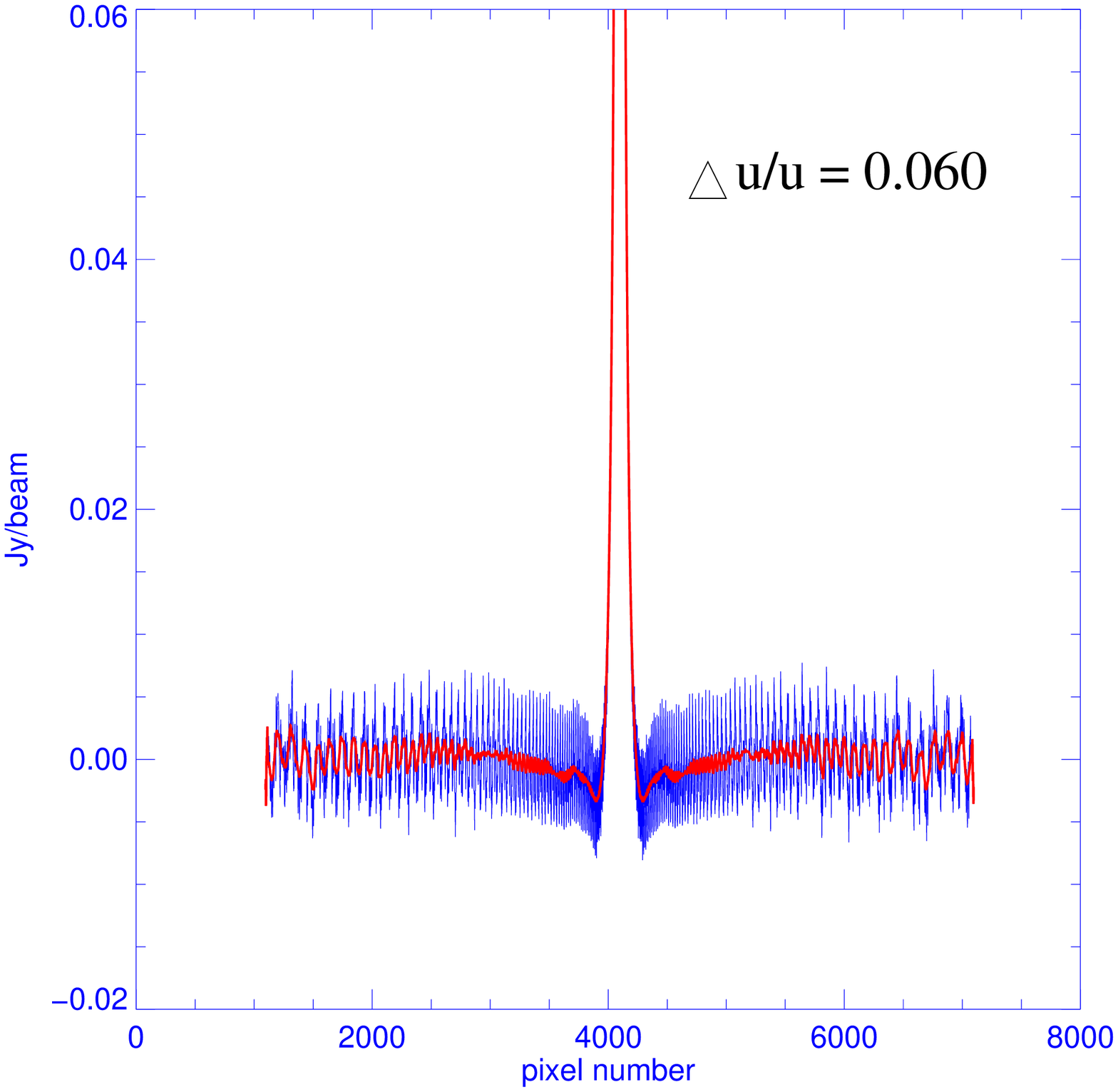} \\ [-0.6cm]
\\
\includegraphics[height=4.7cm]{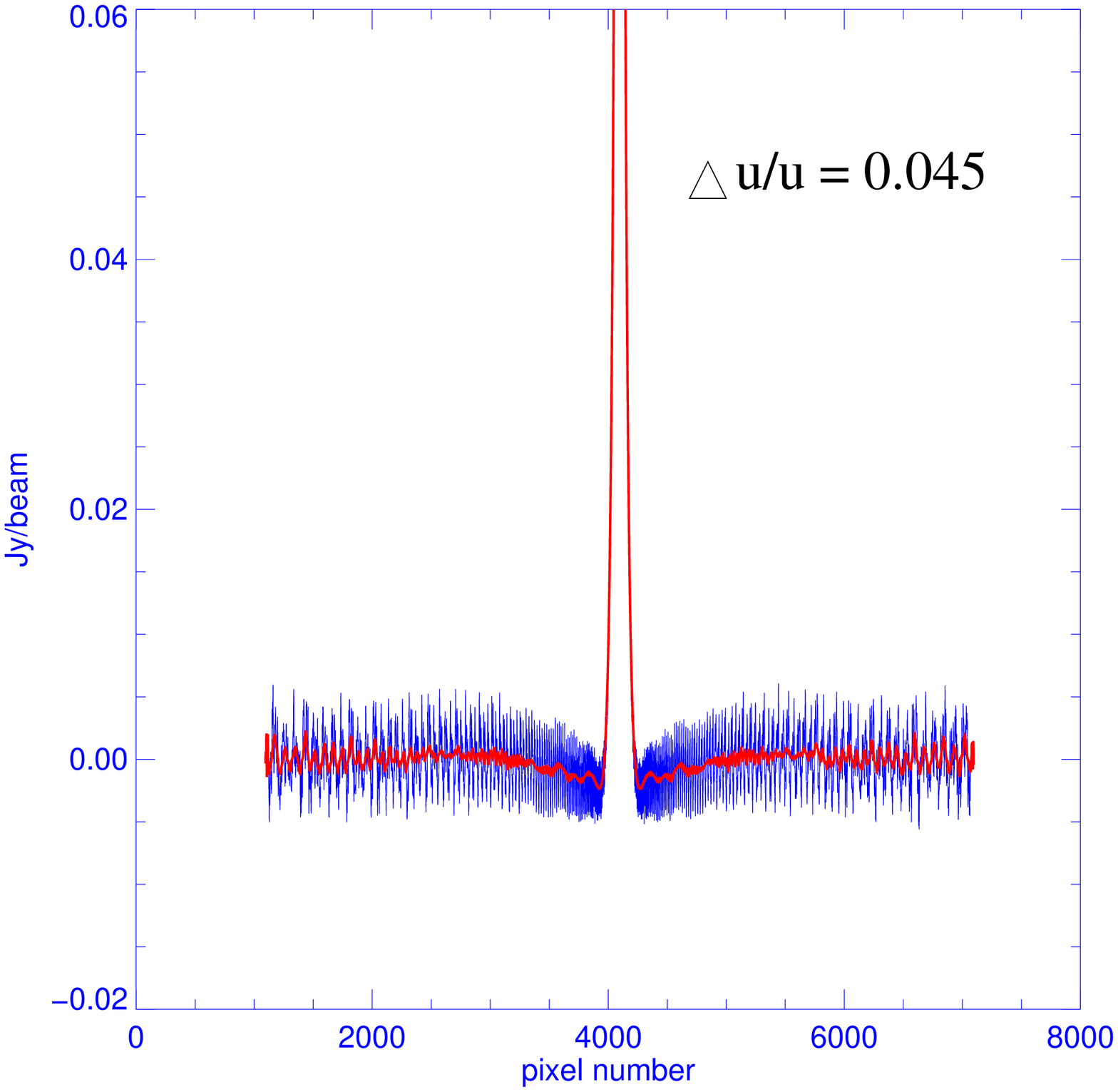} &
\includegraphics[height=4.7cm]{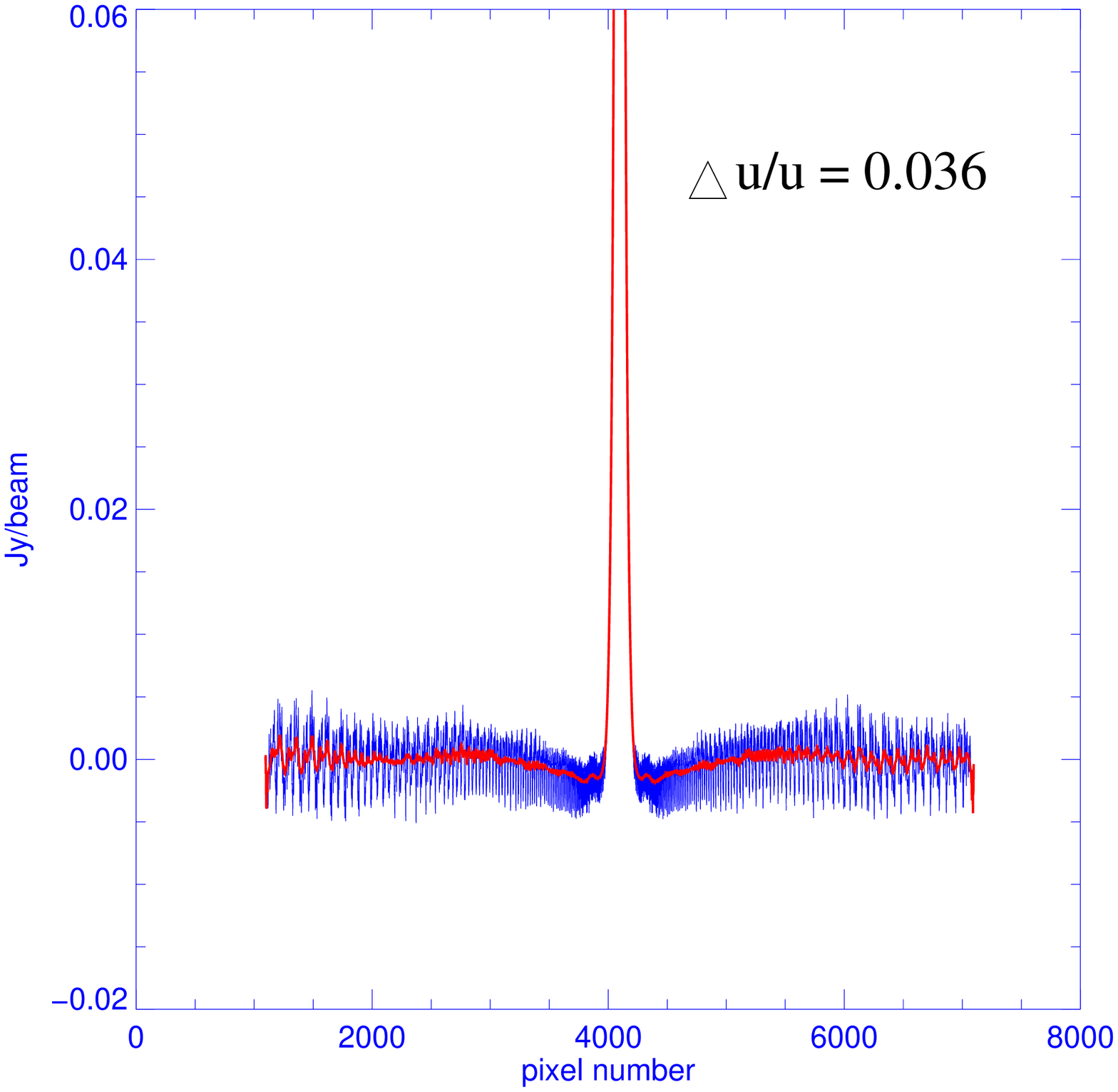} &
\includegraphics[height=4.7cm]{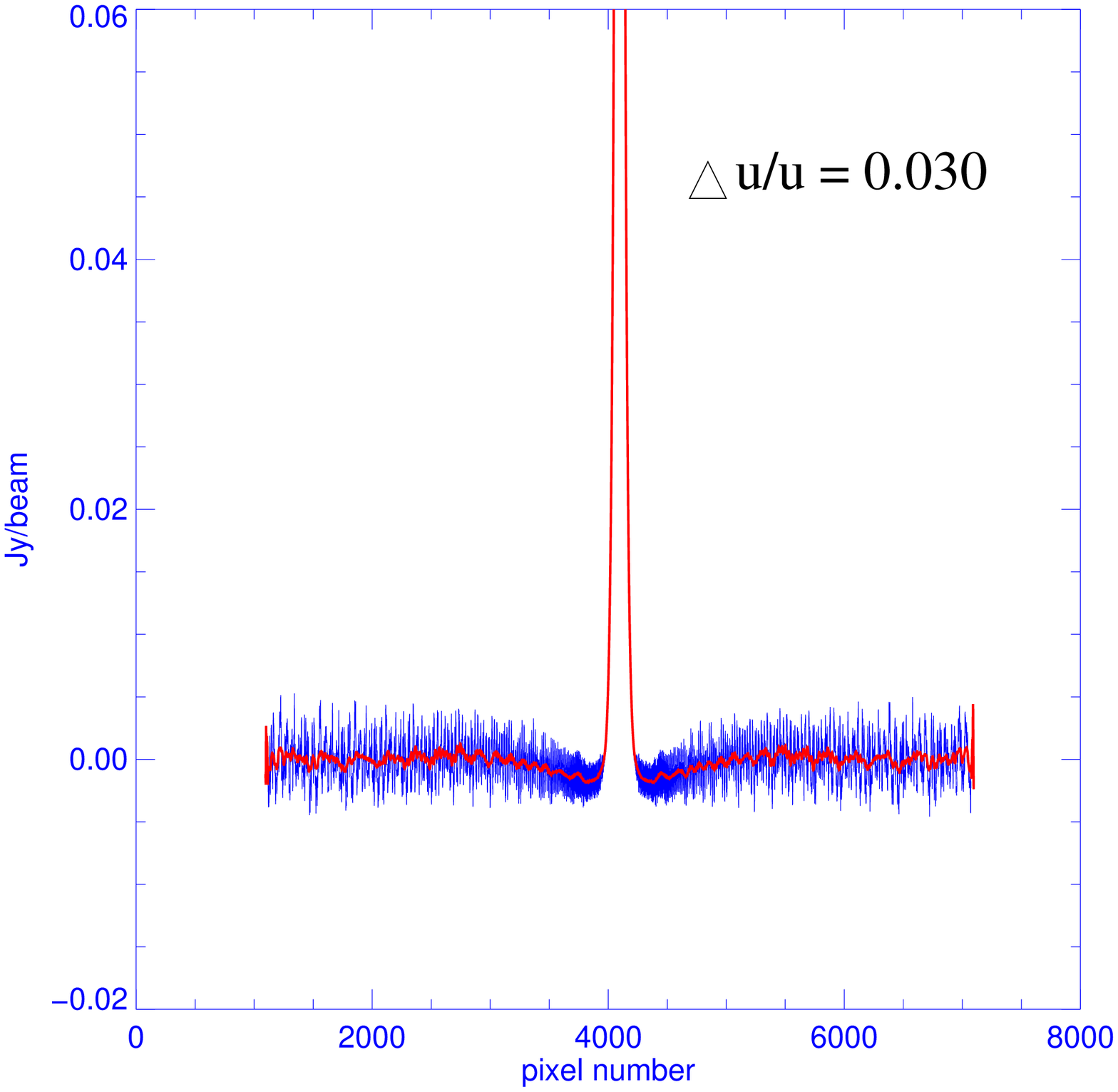} \\ [-0.6cm]
\\
\includegraphics[height=4.7cm]{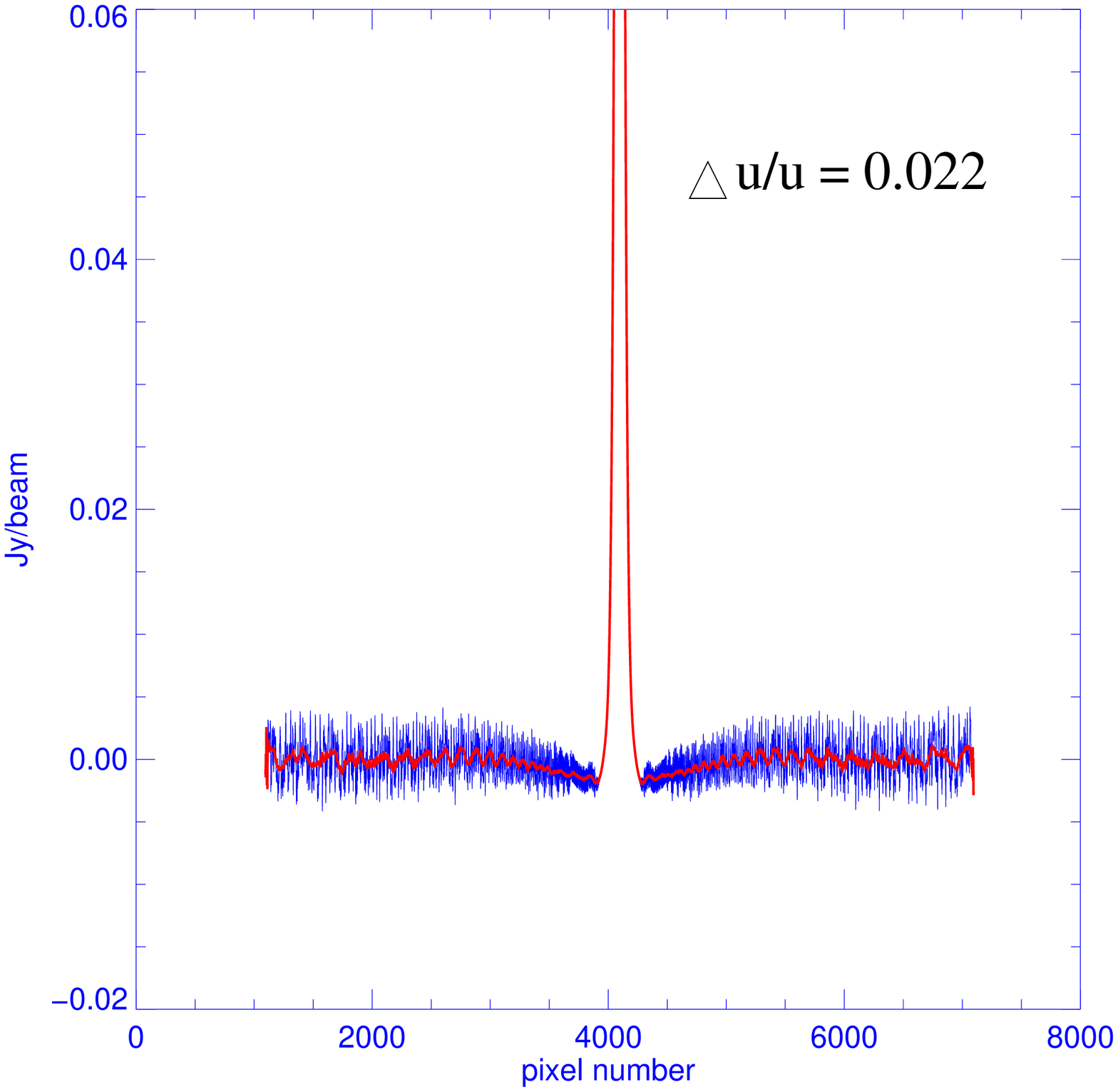} &
\includegraphics[height=4.7cm]{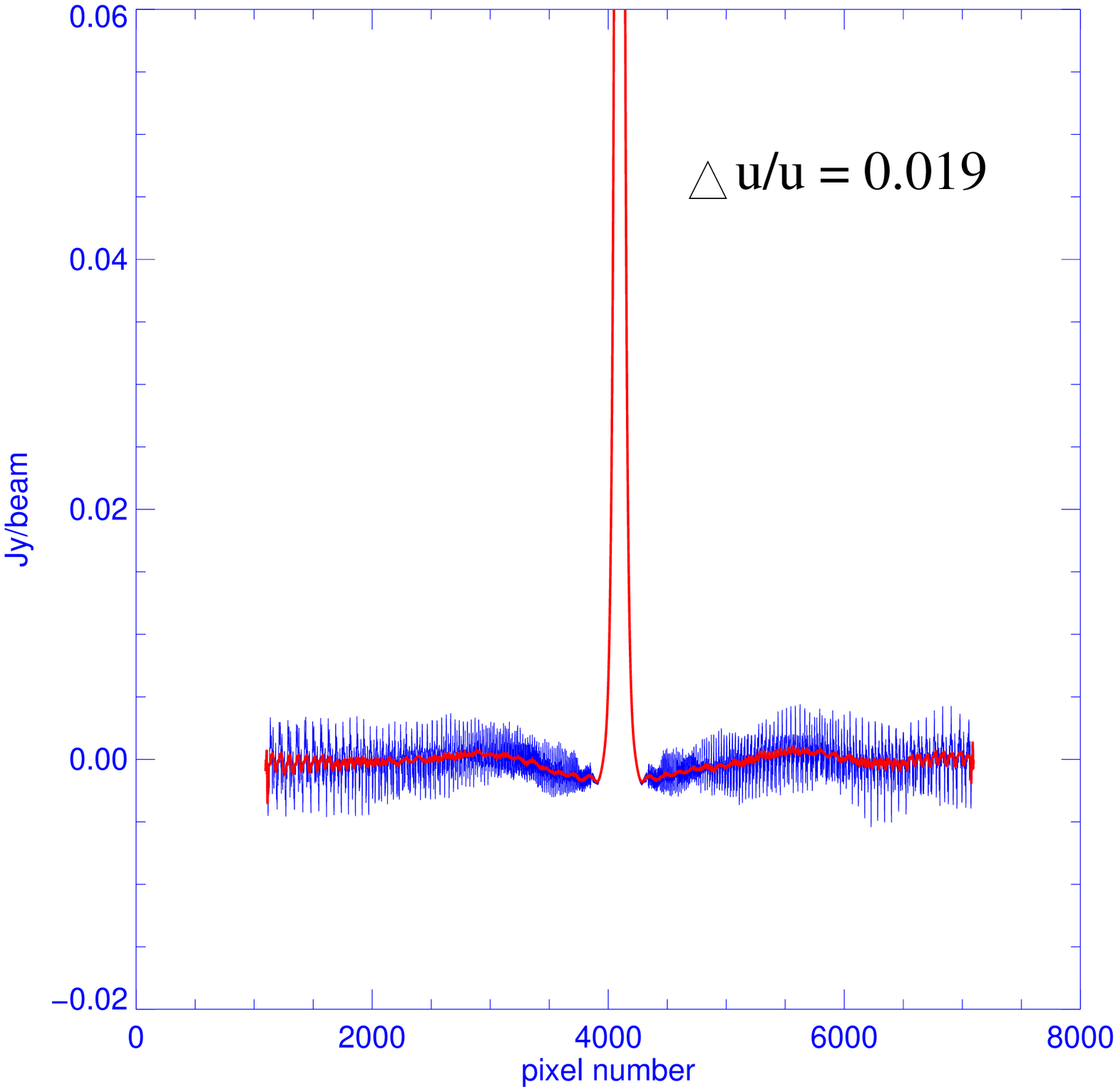} &
\includegraphics[height=4.7cm]{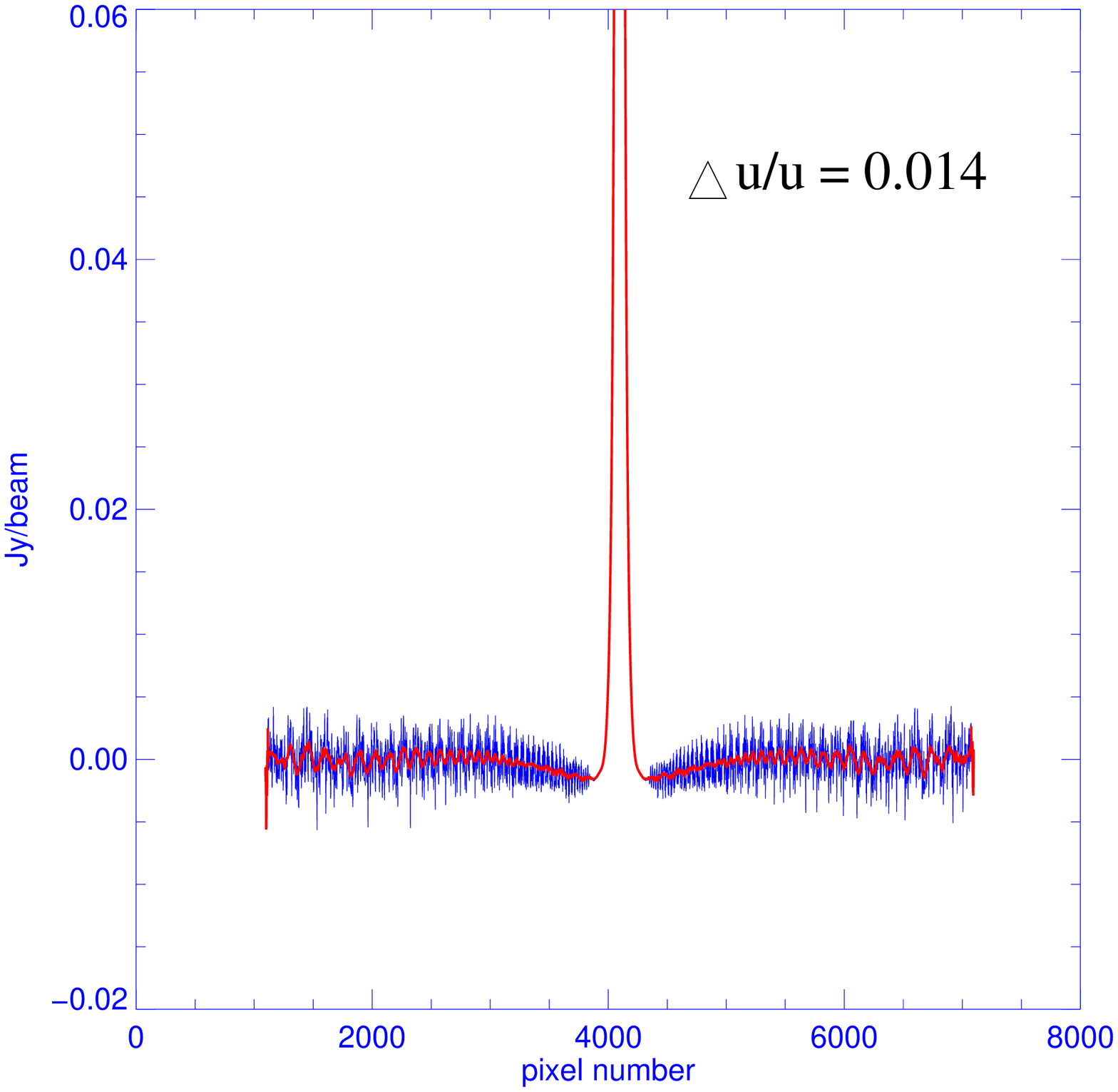} \\ [-0.6cm]
\end{tabular}
\end{center}
\caption{One-dimensional profiles of restoring beams obtained from selected
datasets.
Each panel shows the value of $\Delta u/u$ corresponding to the respective
simulated dataset.
The profiles are truncated at 0.06 of the beam peak in order to
emphasise the difference in the sidelobes resulting from different
{\em uv}-coverages.  In each panel, the red curve represents the
30-pixel boxcar smoothing of the respective beam profile.}
\label{beam_cut}
\end{figure}

We consider only baselines to the station at the origin of the array
and generate a range of visibility datasets to probe the $uv$-gap
parameter from 0.19 ($N$ = 50) to 0.01 ($N$ = 640). The simulated
intergation time is set to 1 second, and the resulting sampling times
range from 12.5 seconds (one {\em uv}-point every 12.5 seconds) for
$N=50$ to 160 seconds for $N=640$. With these settings, we produce
visibility datasets for full track {\em uv}-coverages for each of the
array configuration.

>From the simulated data, we produce both dirty and CLEAN images, of
8192 pixels $\times$ 8192 pixels in size, with a pixel size of
3~arcsec.  The dirty images obtained for several different simulated
configurations are shown in Fig.~\ref{map12}.  The respective
interferometer beam cuts are shown in Fig.~\ref{beam_cut}.  The CLEAN
images were obtained using {\sc AIPS} task IMAGR applied to 499 facets
spread across a $\sim 50$ square degree field.  It should be noted
that the CLEAN algorithm has been applied non-interactively 
 ({\em
i.e.}, with the same set of CLEAN parameters for all datasets and
without varying them during the CLEAN-ing),
 and only a fraction of the
total flux density has been recovered after 200,000 iterations, which
points out to likely limitations due to application of gridding and
deconvolution.

\begin{figure}
\begin{center}
\begin{tabular}{l}
\includegraphics[height=7.7cm]{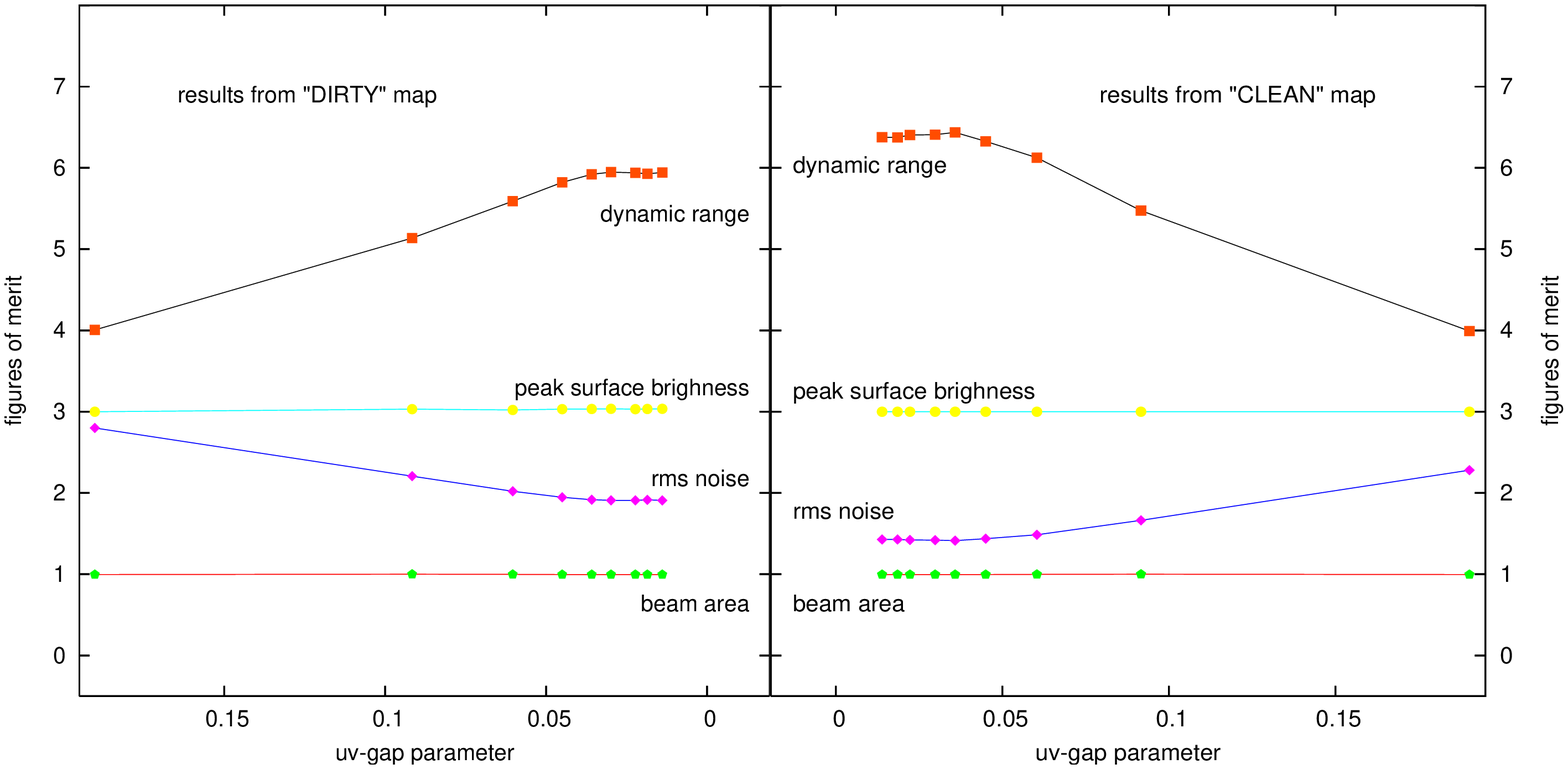}
\end{tabular}
\end{center}
\caption{Relative changes of different FoMs obtained from the dirty
(left anel) and CLEAN (right anel) maps. The normalization is done
with respect to the FoM values determined from the dirty image with
$\Delta u/u =0.01$, by normalizing its beam area to unity, rms noise
to the value of 2, peak surface brightness to the value of 3, and
dynamic range to the value of 6.  The peak surface brightness
recovered from different simulation runs remains nearly constant. The
rms noise first decreases rapidly for smaller values of $\Delta u/u$,
but the rate of this decrease is then remarkably reduced, implying
that the {\em uv}-coverage does not affect strongly the image noise
and dynamic range for array configurations with $\Delta u/u <
0.03$. Note that the dynamic range is determined for a feature with
the size substantially smaller than the largest angular scale for the
{\em uv}-coverage with $\Delta u/u=0.19$, and thus the calculations
are not affected by variable minimum {\em uv}-spacing in different
simulation runs. }
\label{ex_2}
\end{figure}

The resulting facet images were stitched together using {\sc AIPS} task
FLATN to create a single final image.  We used uniform weighting and
the 3D option for the $w$-term correction throughout our analysis
(Cornwell \& Perley 1992).  In order to assess the effect of
deconvolution and gridding of {\em uv}-data on the results of the
simulations, both the CLEAN and the dirty images were used for
estimating several basic FoMs. We use {\sc AIPS} task IMSTAT to determine
the r.m.s. noise levels in each case.  The results of these estimates
are compared in Fig.~\ref{ex_2}.

One can see that the dependence of the rms noise (and reciprocally,
the dynamic range) on the {\em uv}-coverage changes at $\Delta u/u
\approx 0.03$. At smaller values of $\Delta u/u$, the {\em
uv}-coverage does not have a strong effect on the flux density
recovered from the visibility data (this holds true for both dirty and
CLEAN images). The requirement of 
\[
\Delta u/u \lesssim 0.03
\]
for the entire range of baselines can therefore be used as a benchmark
requirement for designing the SKA configurations that would minimise
the effect of {\em uv}-coverage on reconstructing the sky brightness
distribution.  
It should be noted that this conclusion
provides a strong benchmark, implying implementing {\em uv}-coverages
with {\em uv}-gaps larger than 0.03 will limit the imaging capabilities of an array {\em even with although the present day convolution and imaging 
algorithms.}

Further, more detailed investigations may be required in order to
refine this conclusion and assess the full range of effects that may
potentially affect the dynamic range and structural sensitivity
derived from images obtained with different {\em uv}-coverages and
different values of $\Delta u/u$. One possibility would be to use more
elaborate and more realistis sky models coming from source simulations
({\em e.g.}, O'Sullivan et al. 2009).

\section{Practical issues}

We illustrate now a practical way to use the $\Delta u/u$ parameter
for evaluating an arbitrary array configuration. The evaluation is
based on analysis of a {\em uv}-coverage arising from a specific
observation or a set of observations (for instance, a number of
snapshots on targets covering a specified range of declinations and/or
hour angles).

Since almost all real {\em uv}-coverages deviate from
circular symmetry, the $\Delta u/u$ FoM  should
be represented either by a two-dimensional distribution or by an average of
that. The averaging can be made azimuthally (providing a profile of
$\Delta u/u$ as a function of {\em uv}-distance) or both azimuthally and
radially (giving a single value description of a given {\em
uv}-coverage). For the averaged quantities, the respective
dispersions can be used to quantify inhomogeneities in the {\em
uv}-coverages.

For the purpose of a better graphical representation of the {\em
uv}-gap distribution, a definition $\Delta u/u = (u_2 - u_1)/u_2$
($u_2\ge u_1$) can be applied. This definition changes the range of
$\Delta u/u$ from $[0,\infty]$ to $[0,1]$, with 0 corresponding to
$\Delta u/u$ from two identical baselines. The maximum value $\Delta
u/u = 1$ is realized everywhere outside the area covered by the
observation (for which $u_2 = \infty$ can be assumed).

A feasible approach to determine the figures of merit based on $\Delta u / u$
is as follows:
\begin{enumerate}
\item Grid the {\em uv}-data into $N$ sectors, where the width of each sector is determined by the observing scan-length (the duration of a snapshot observation, or a typical length of a single scan in a synthesis observation).
\item For each individual sector (described by its central position angle, $\phi_i$ and
width $2\pi/N$ and containing $M$ {\em uv}-points), determine $\Delta u/u (u,\phi_i)$ for all $M$ {\em uv}-points. As was pointed above, $\Delta u/u(u,\phi_i) < 1$ within the range  $u_1 \le u \le u_M$ and it is equal to unity elsewhere.
\item Plot the combined results for all sectors in polar or
rectangular coordinates or
perform averaging in azimuth $\langle
\frac{\Delta u}{u} (u, \phi) \rangle_{\phi}$ and/or azimuth and
radius $\langle \frac{\Delta u}{u} (u, \phi) \rangle_{u,\phi}$.
\end{enumerate}

A detailed description of the calculaton algorithm is given in
Appendix~1.  Appendix~2 summarizes a proposed set of FoM that includes
the {\em uv}-gap parameter and can be used for evaluation and
optimization of the SKA configuration.

\subsection{Examples}

In order to represent a two-dimensional distribution $\Delta u/u$, the
Voronoi tessellation can be applied to combine the $\Delta u/u$
calculated for the individual sectors. The Voronoi tessellation
("Voronoi diagram") uses partitioning of a plane with $n$ points
into convex polygons such that each polygon contains exactly one
generating point and every point in a given polygon is closer to its
generating point than to any other (Okabe et al. 2000).  The Voronoi
diagram is sometimes also known as a Dirichlet tessellation. The cells
are called Dirichlet regions, Thiessen polytopes, or Voronoi polygons.
We use the algorithm by Okabe et al. (2000) built in {\sc MATLAB} and GNU
Octave (version 3.2) to decompose the discrete set of $\frac{\Delta
u}{u} (uv, \phi_i$ ($i=1,N$) values into density plots that can be
used as a graphical representation of the {\em uv}-gap distribution.

\begin{figure}
\begin{center}
\begin{tabular}{ll}
\includegraphics[height=6.3cm]{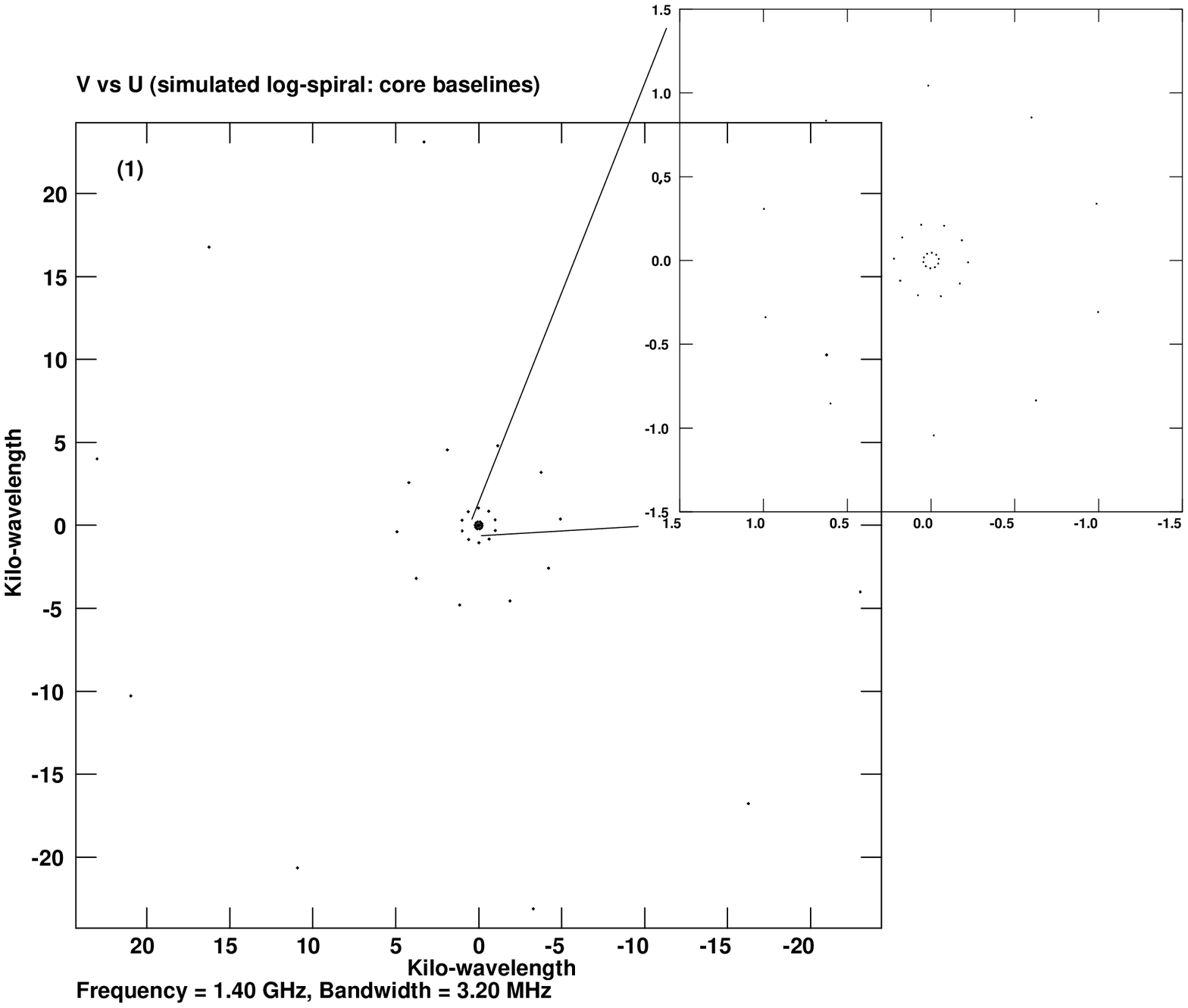} &
\includegraphics[height=6.3cm]{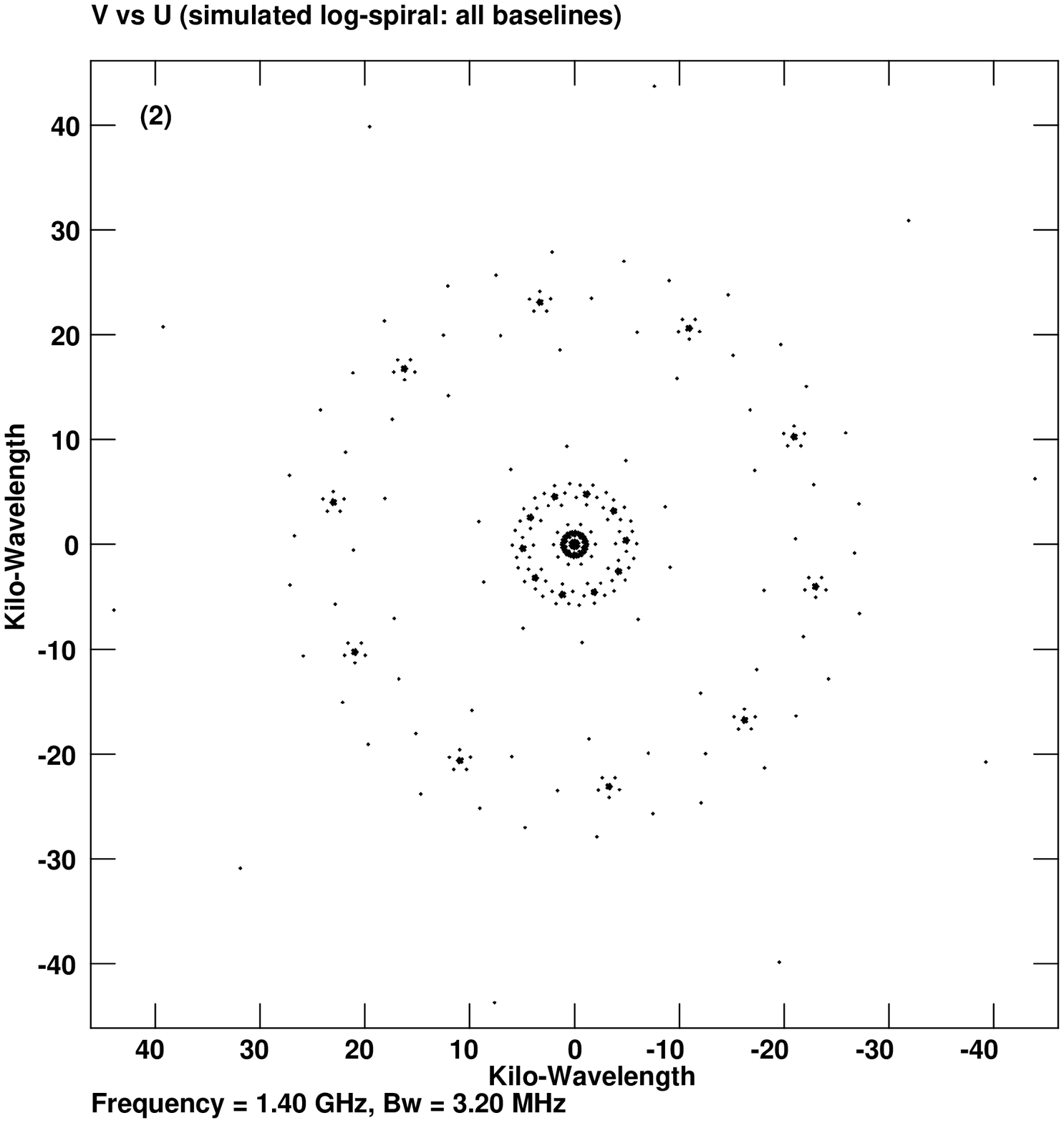} \\ [-0.6cm]
\\ 
\includegraphics[height=6.3cm]{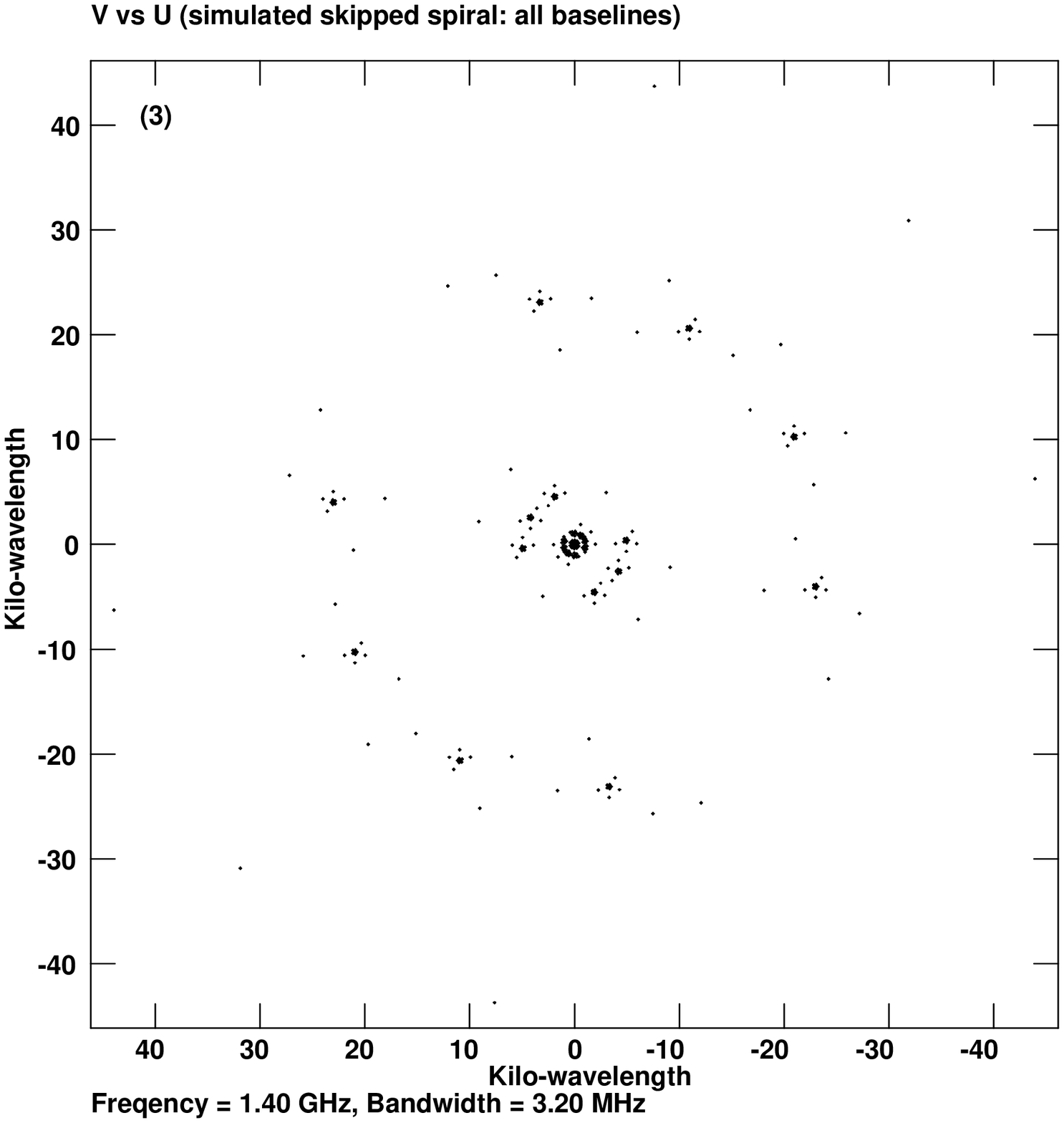} &
\includegraphics[height=6.3cm]{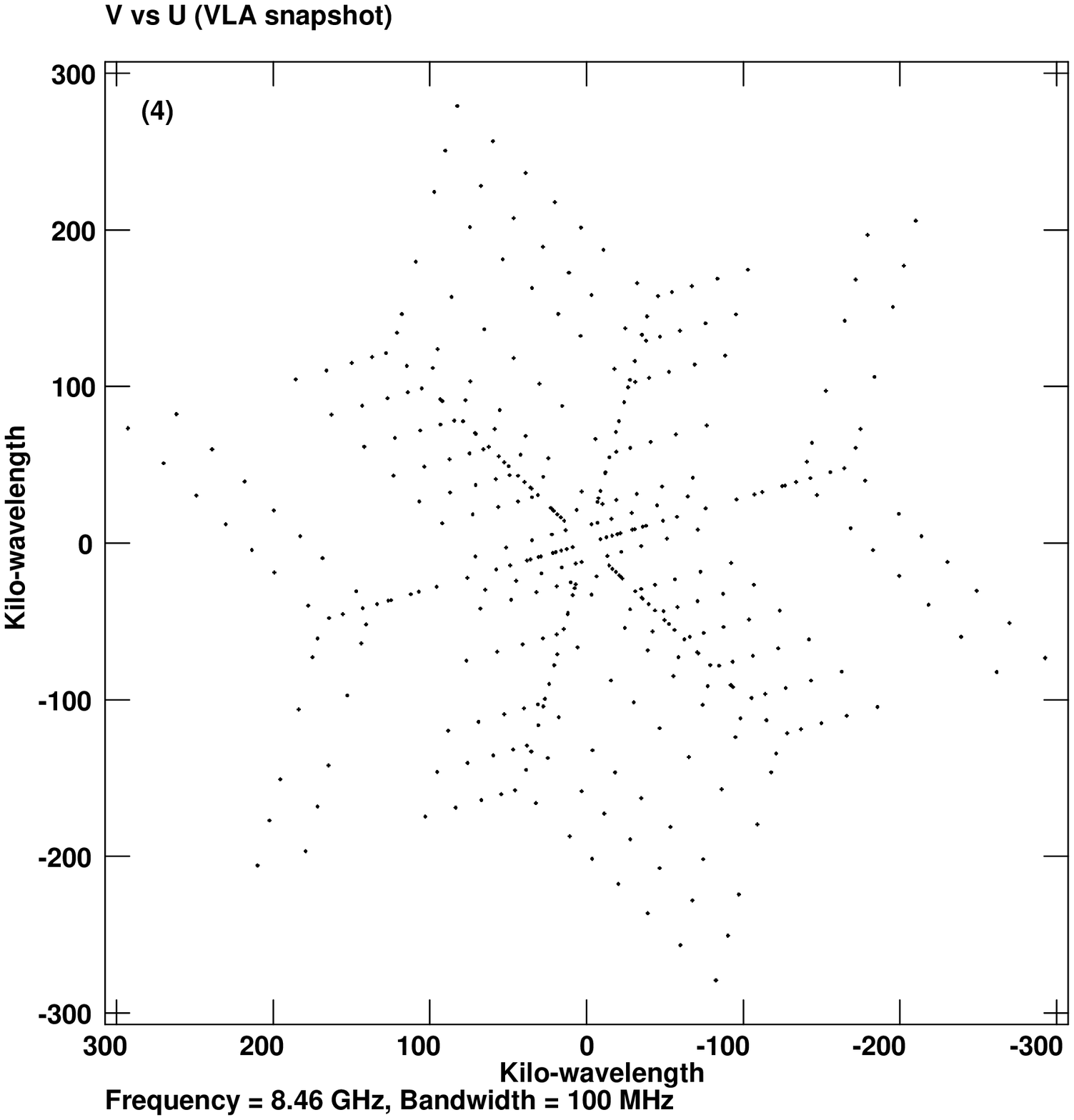} \\ [-0.6cm]
\\ 
\multicolumn{2}{c}{\includegraphics[height=6.3cm]{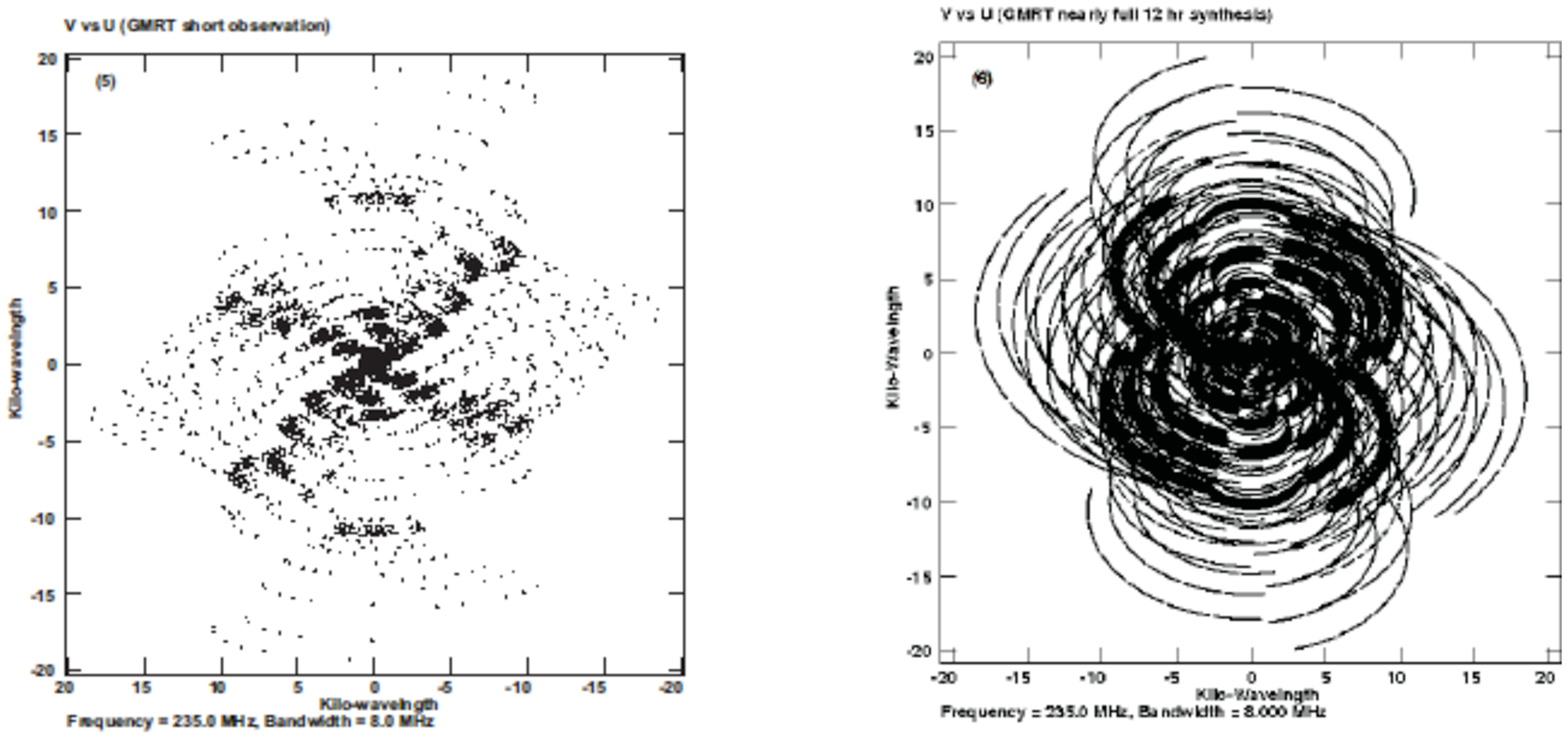}} \\ [-0.4cm]
\end{tabular}
\end{center}
\caption{ Six different {\em uv}-coverages used to exemplify
determination of the $\Delta u / u$ figures of merit: (1)~a {\em
uv}-coverage from a simulated log-spiral, baselines to the core
station included; (2)~the same simulation, all baselines included;
(3)~a {\em uv}-coverage from a simulated skipped-spiral configuration
(aimed to provide a nearly constant $\Delta u/u$ and to satisfy the
collecting area distribution required for the SKA); (4)~a VLA snapshot
{\em uv}-coverage; (5)~a GMRT {\em uv}-coverage from a short
observation; (6)~a GMRT {\em uv}-coverage from a nearly complete,
12-hour synthesis.}
\label{sim_real}
\end{figure}

\begin{figure}
\begin{center}
\begin{tabular}{ll}
\includegraphics[height=5.2cm]{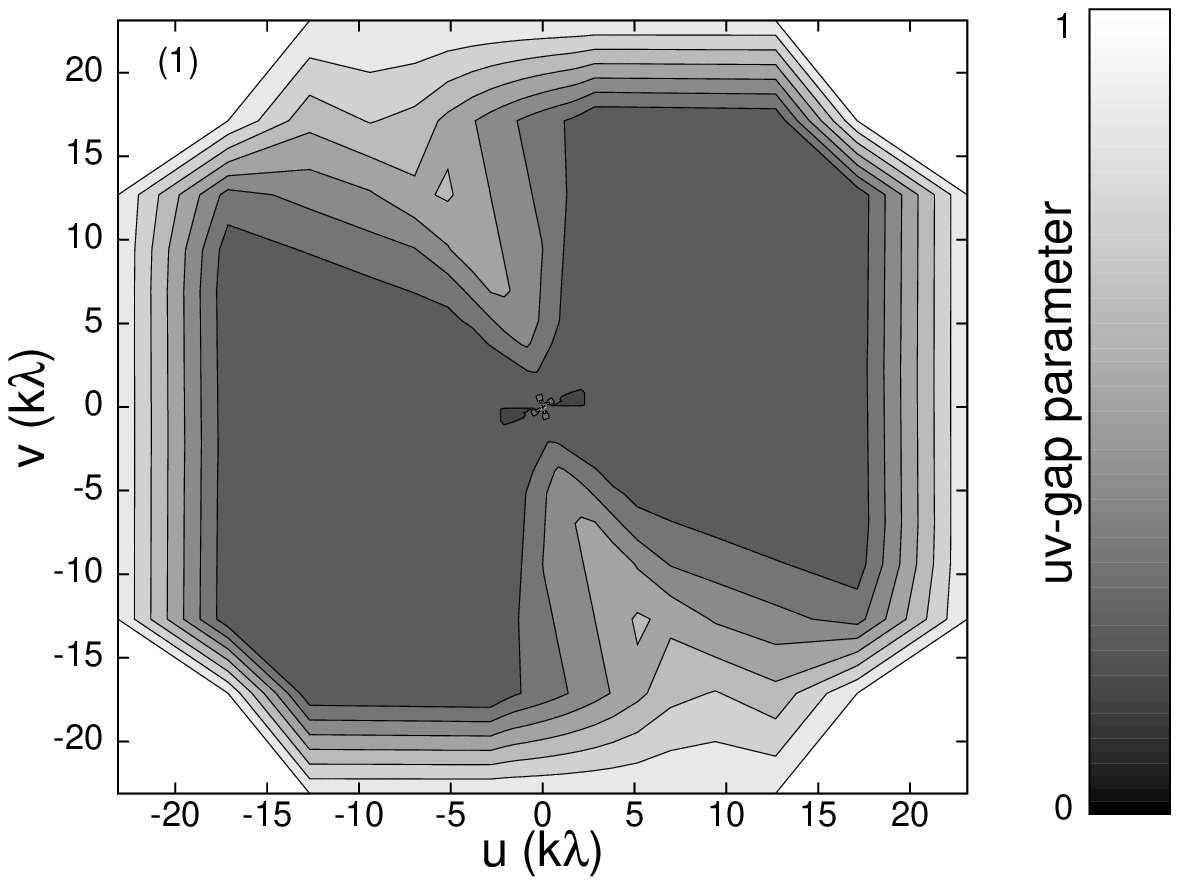} &
\includegraphics[height=5.2cm]{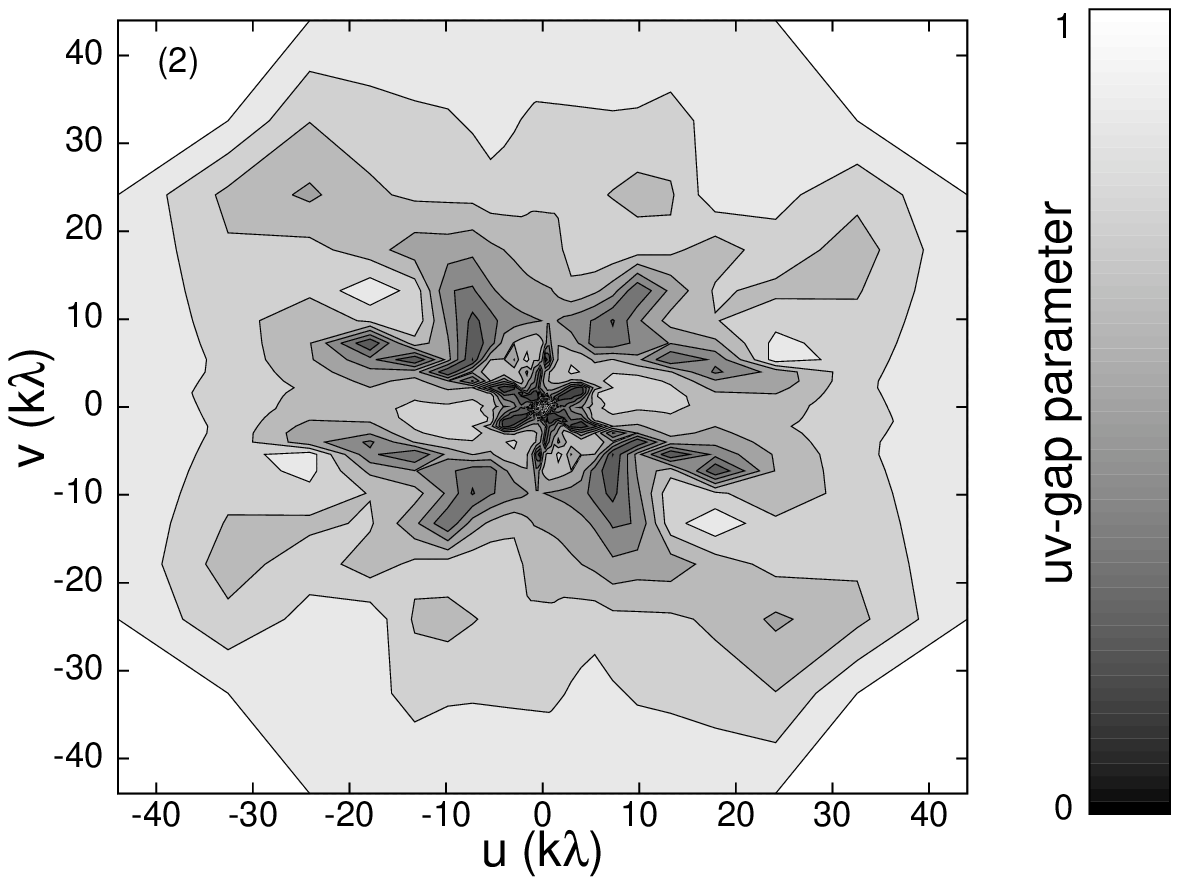} \\
\\ 
\includegraphics[height=5.2cm]{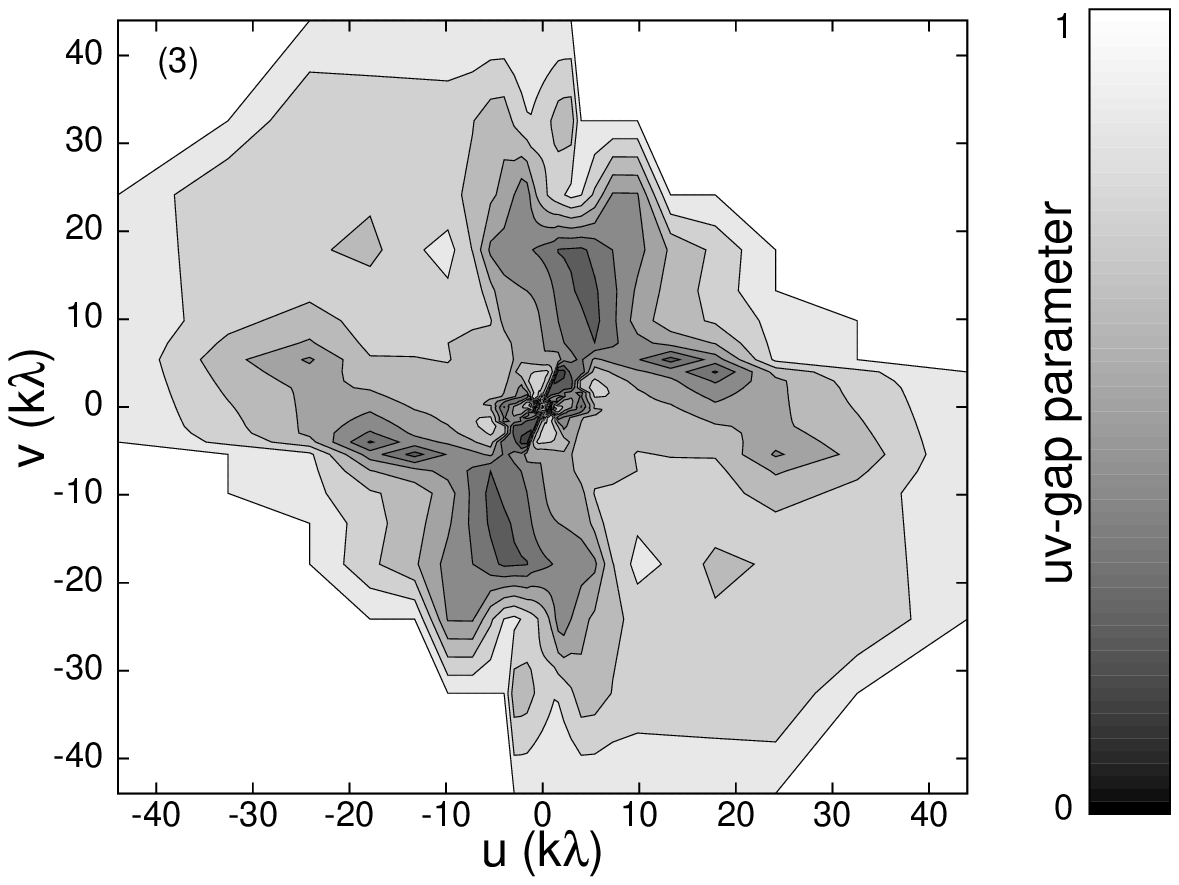} &
\includegraphics[height=5.2cm]{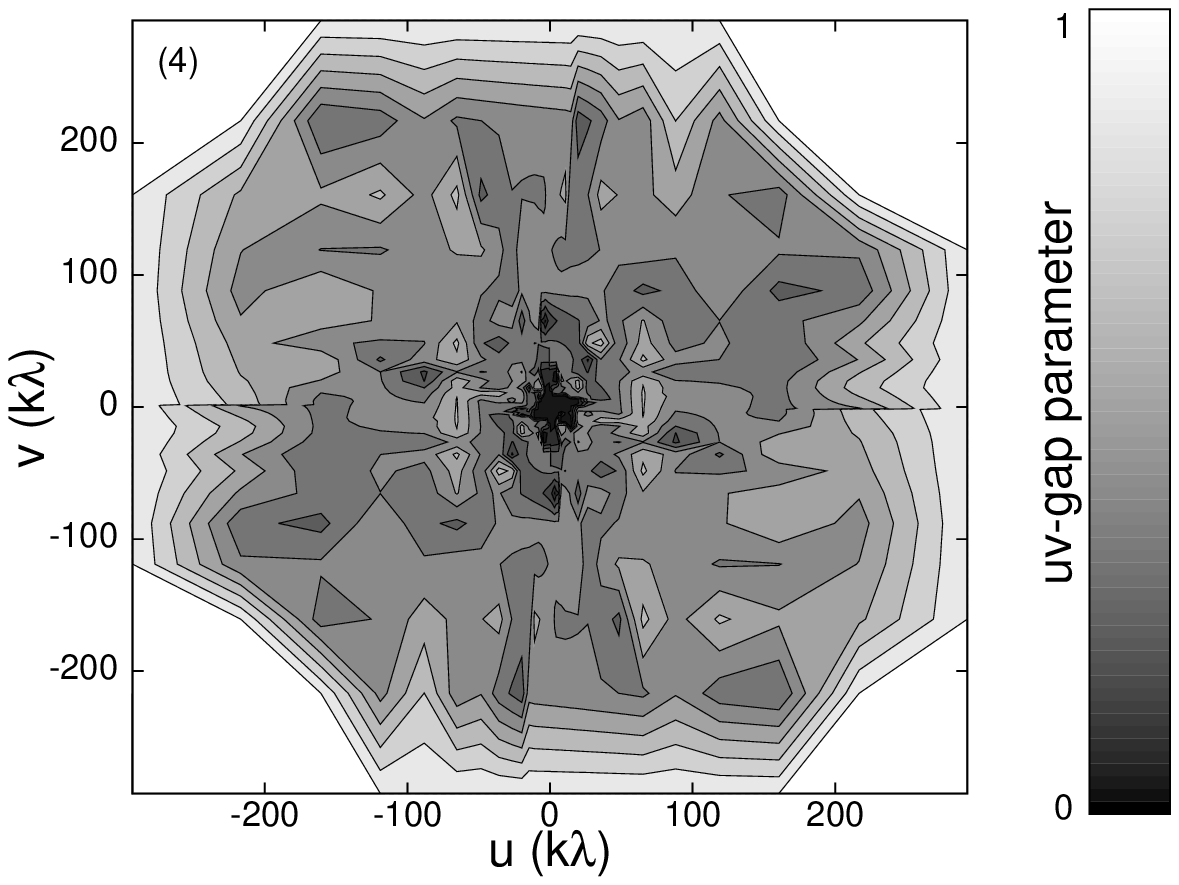} \\
\\
\includegraphics[height=5.2cm]{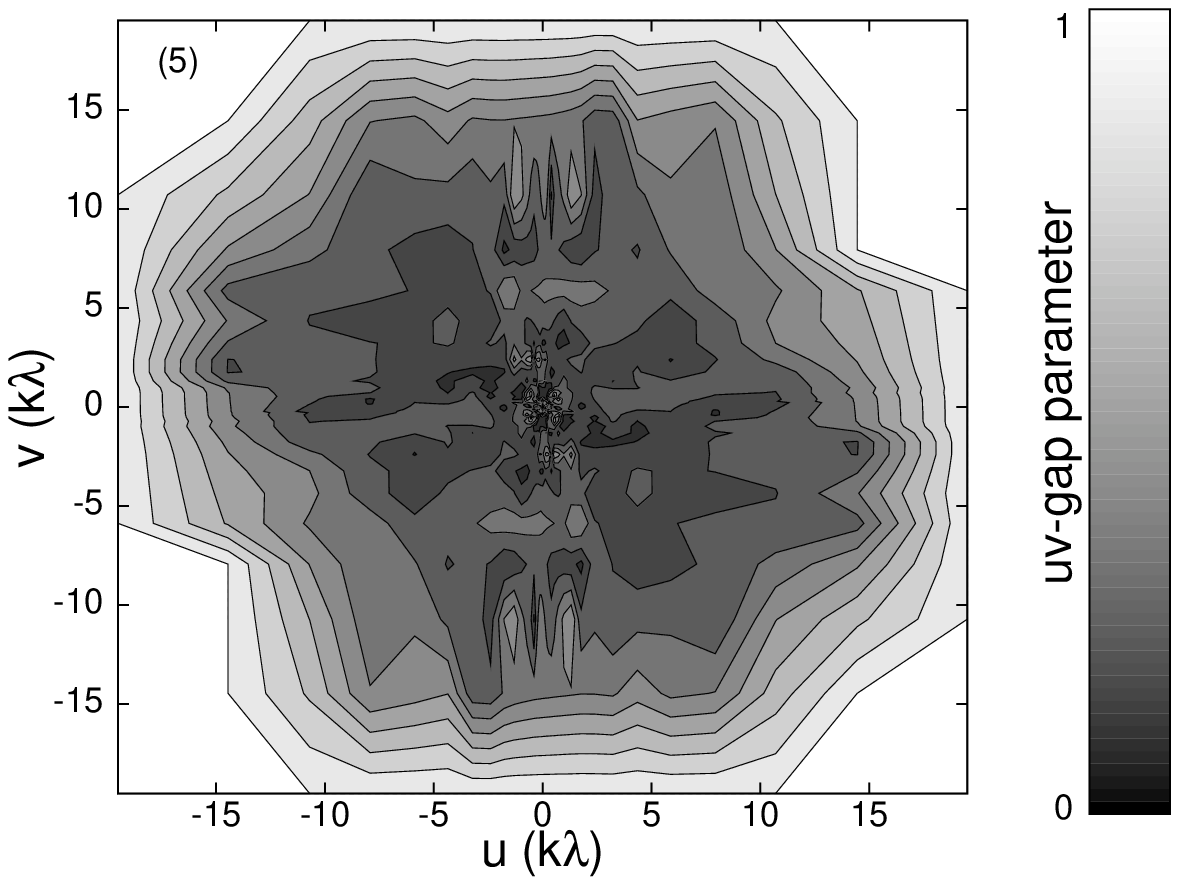} &
\includegraphics[height=5.2cm]{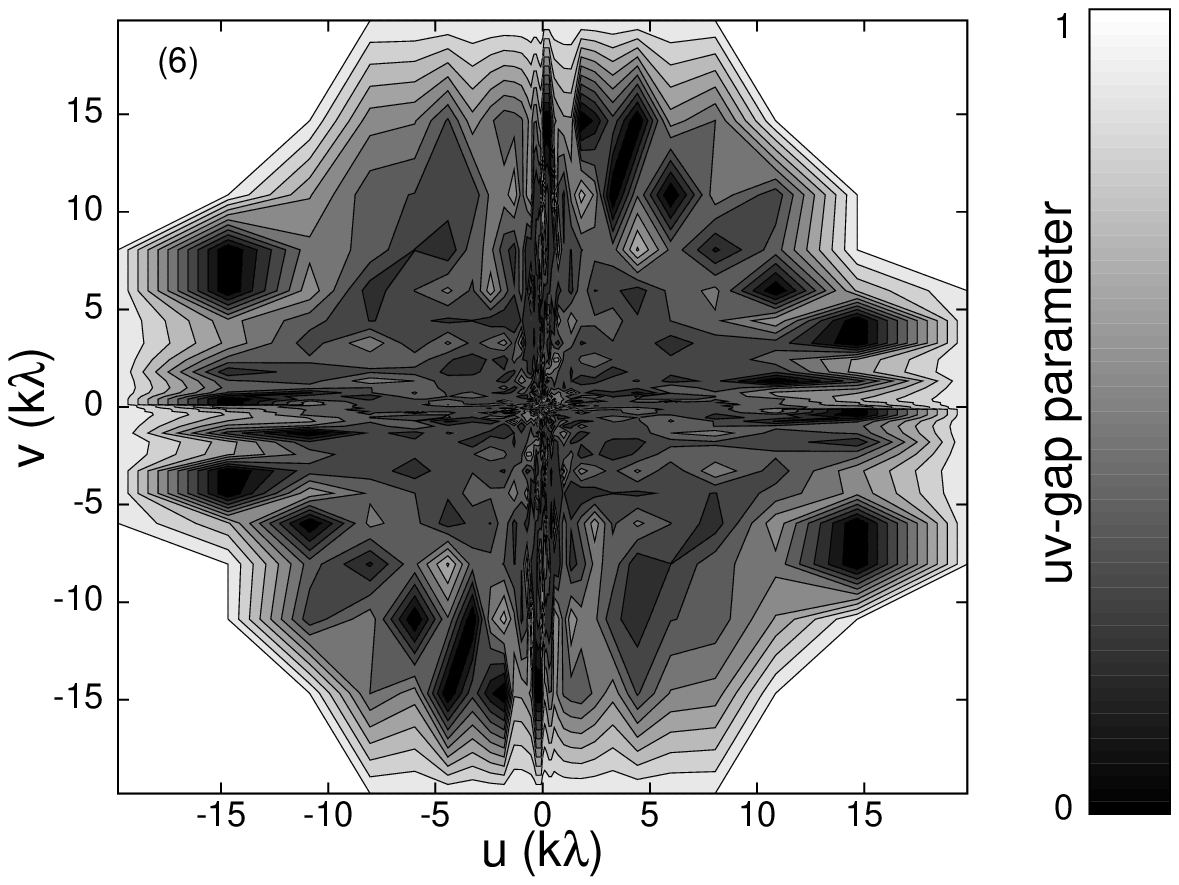}
\end{tabular}
\end{center}
\caption{Density plots of  $\frac{\Delta u}{u} (u, \phi)$
distribution calculated from the {\em uv}-coverages shown in Fig~5. 
Note that for very sparce {\em uv}-coverages, the tesselation algorithm may introduce artefacts. This can be illustrated by the apparent asymmetry of $\Delta u/u$ distribution seen in panel 1 produced for a symmetric {\em uv}-coverage with very few {\em uv}-points at large {\em uv}-radii). More refined approaches to calculating the density fields from {\em uv}-coverages may be needed for such cases.
}
\label{density_plot}
\end{figure}

We exemplify this approach by considering six different {\em
uv}-coverages from simulated and real observations
(Fig.~\ref{sim_real}). The resulting two-dimensional distributions of
$\Delta u/u$ are shown in Fig.~\ref{density_plot}. One can either analyze these distributions directly, or produce azimuthally-averaged radial profiles, or describe these {\em uv}-coverages by a single FOM obtained from double, azimthal-radial averaging. Table~\ref{azimuth_res} shows values of $\Delta u/u$ from
azimuthal-radial averaging and the corresponding dispersions for all
panels with numbers 1 through 6 shown in Fig.~\ref{sim_real}.

The radial profiles obtained by azimuthal averaging of the density
plots are shown in Fig.~7. One can see immediately, from the profiles
in Fig.~7 and the dispersions of the mean in Table~\ref{azimuth_res},
that $\Delta u/u$ varies strongly with the baseline length. This
dependence should be minimised during the design of the SKA
configuration, taking into account the geographical location of the
array, the different types of observations ({\em e.g.,} snapshots,
full track synthesis and so on), and the range of declinations for
which the imaging performance of the SKA should be optimised. 

\begin{table}
\caption{Table showing values of $\Delta u/u$ from azimuthal-radial averaging
and the corresponding dispersions for all panels with numbers 1 through 6
in Figs. 6 and 7.}
\centering
\begin{tabular}{ll|cc}
\hline
 & & & \\
 & Example $uv$-coverage & $\langle\Delta u/u\rangle$ & $\sigma_{\Delta u/u}$ \\
 & & & \\
\hline
 & & & \\
(1) & Simulated log-spiral: core baselines           & 0.824 & 0.071 \\
(2) & Simulated log-spiral: all baselines                           & 0.219 & 0.244 \\
(3) & Simulated skipped spiral$^\dag$: all baselines& 0.203 & 0.246 \\
(4) & VLA snapshot            & 0.072 & 0.084 \\
(5) & GMRT short observation            & 0.016 & 0.043 \\
(6) & GMRT nearly full 12 hr synthesis & 0.012 & 0.035 \\
 & & & \\
\hline
\end{tabular}
\\{\small Note: $\dag$ -- a multi-arm, logartihmic spiral configuration, with
concurrent antennas removed from all but one of the arms of the spiral, in
order to reduce the number of redundant baselines as much as possible.}
\label{azimuth_res}
\end{table}

\section*{Acknowledgements}
This effort/activity is supported by the
European Community Framework Programme 6, Square Kilometre Array Design Studies
(SKADS), contract no 011938.
SJM has been partially supported by the Spanish DGYCIT grant
AYA2006-14986-CO2-02.
DVL thanks N. Roy for giving classes on (equiangular) log-spiral,
M.A. Voronkov and R.V. Urvashi for helping with
the aips$++$, and is grateful to K. Borkowski and T. Fukushima. for sharing
their "Cartesian to geodetic coordinates transform" subroutines.
DVL also thanks A.L. Roy
for several fruitful discussions and suggestions. We thank the referee, Craig Walker for valuable comments and suggestions.

\begin{figure}
\begin{center}
\begin{tabular}{ll}
\includegraphics[height=6.0cm]{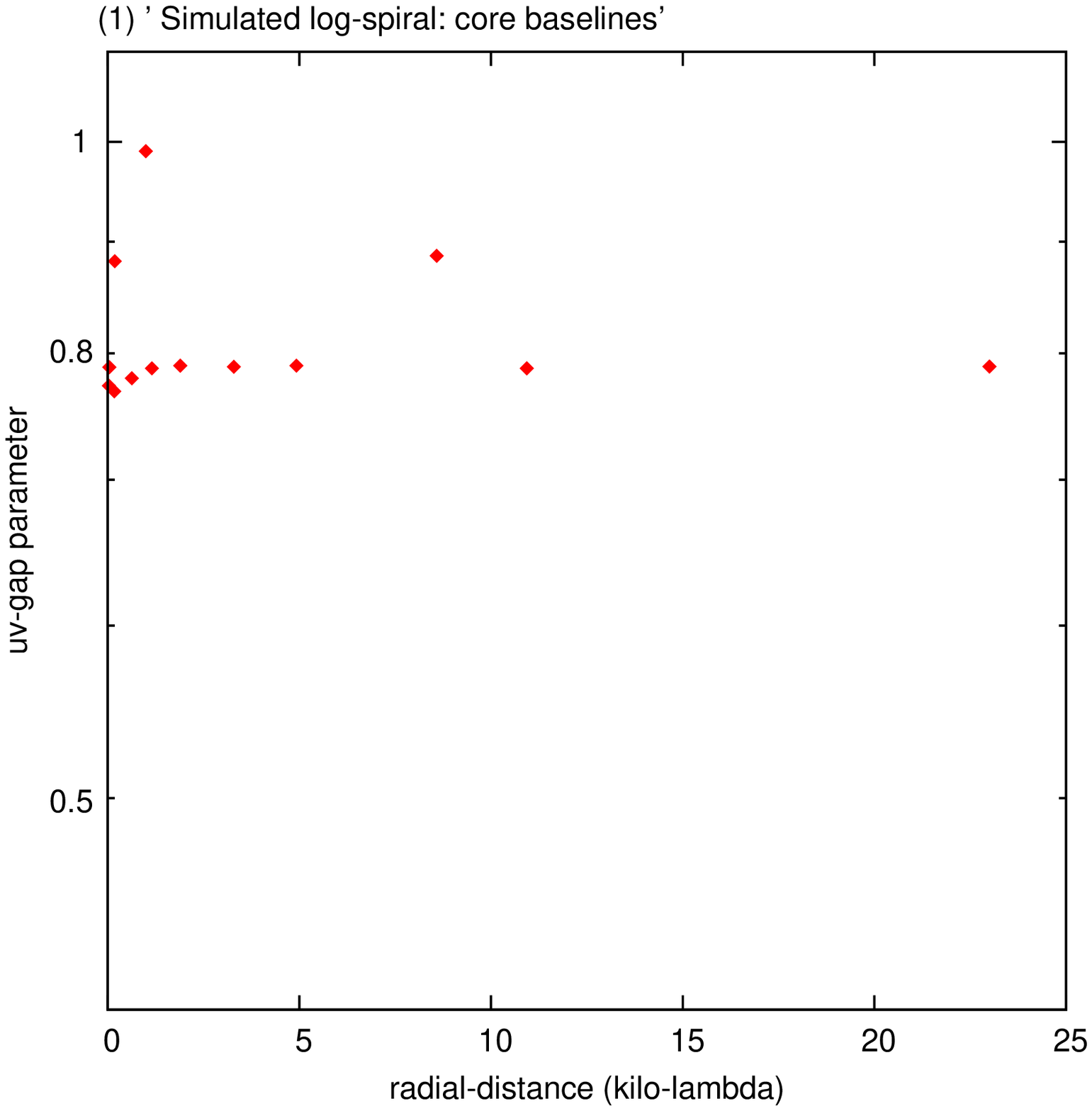} &
\includegraphics[height=5.94cm]{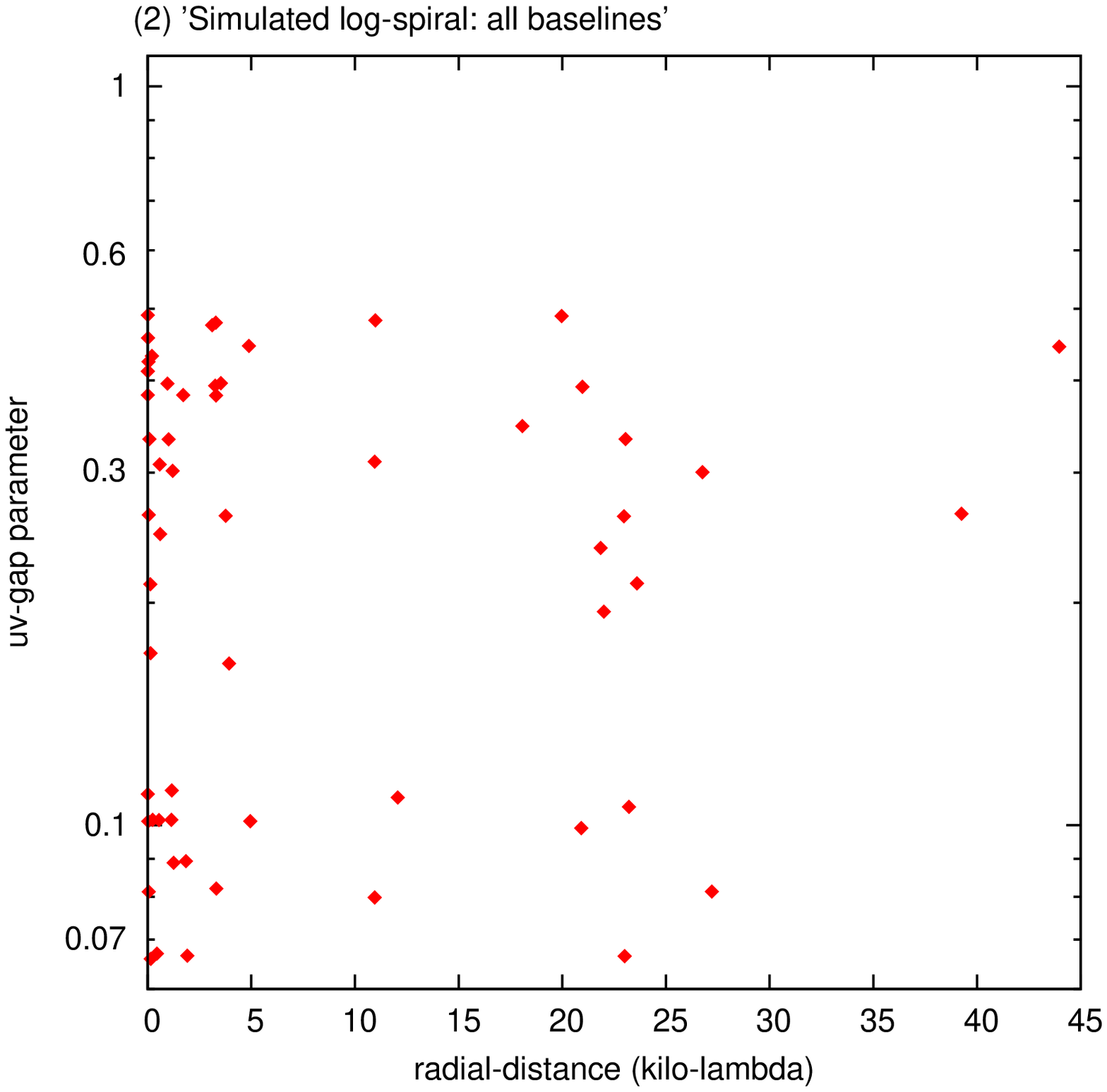} \\
\\
\includegraphics[height=5.94cm]{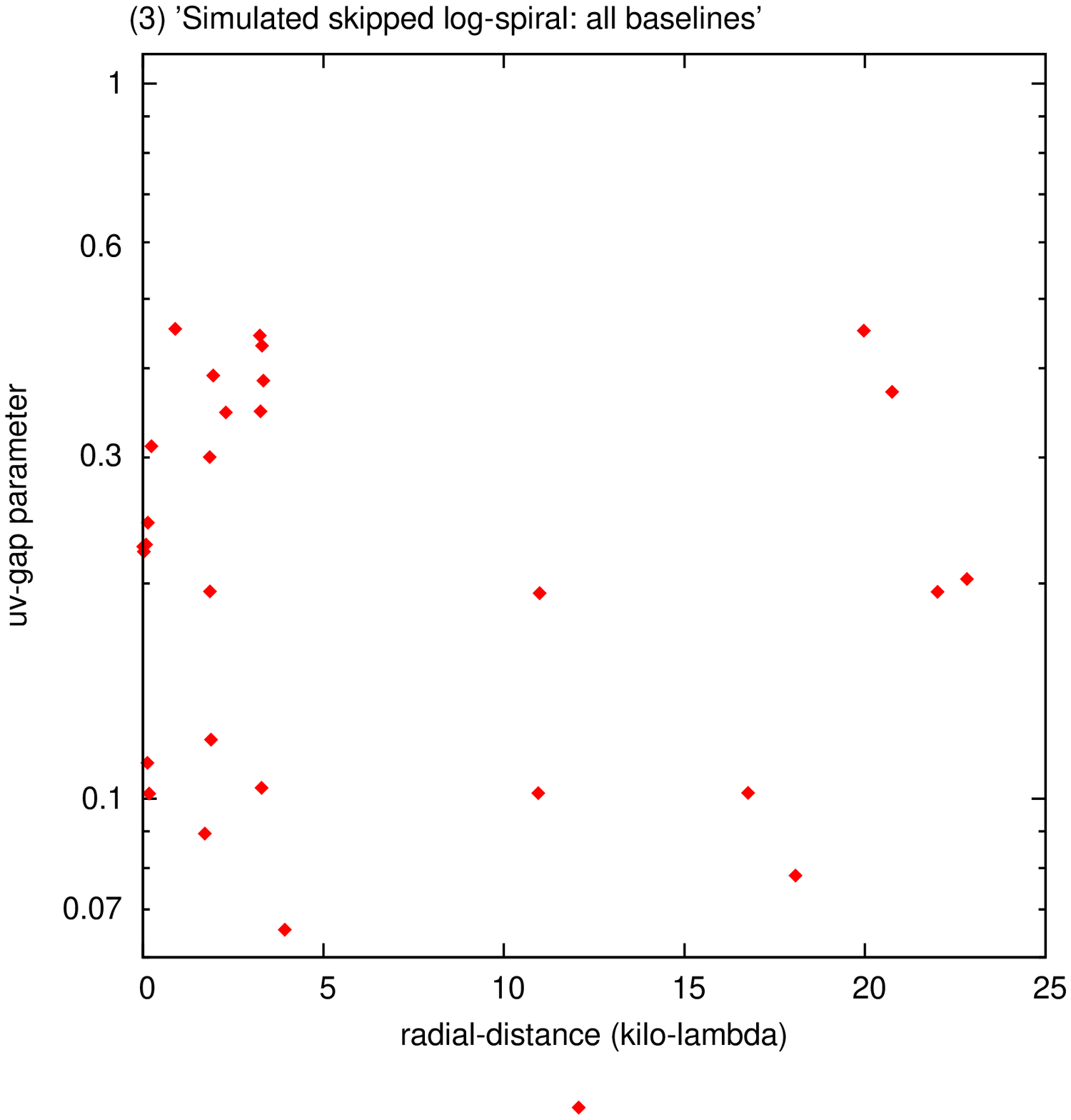} &
\includegraphics[height=5.58cm]{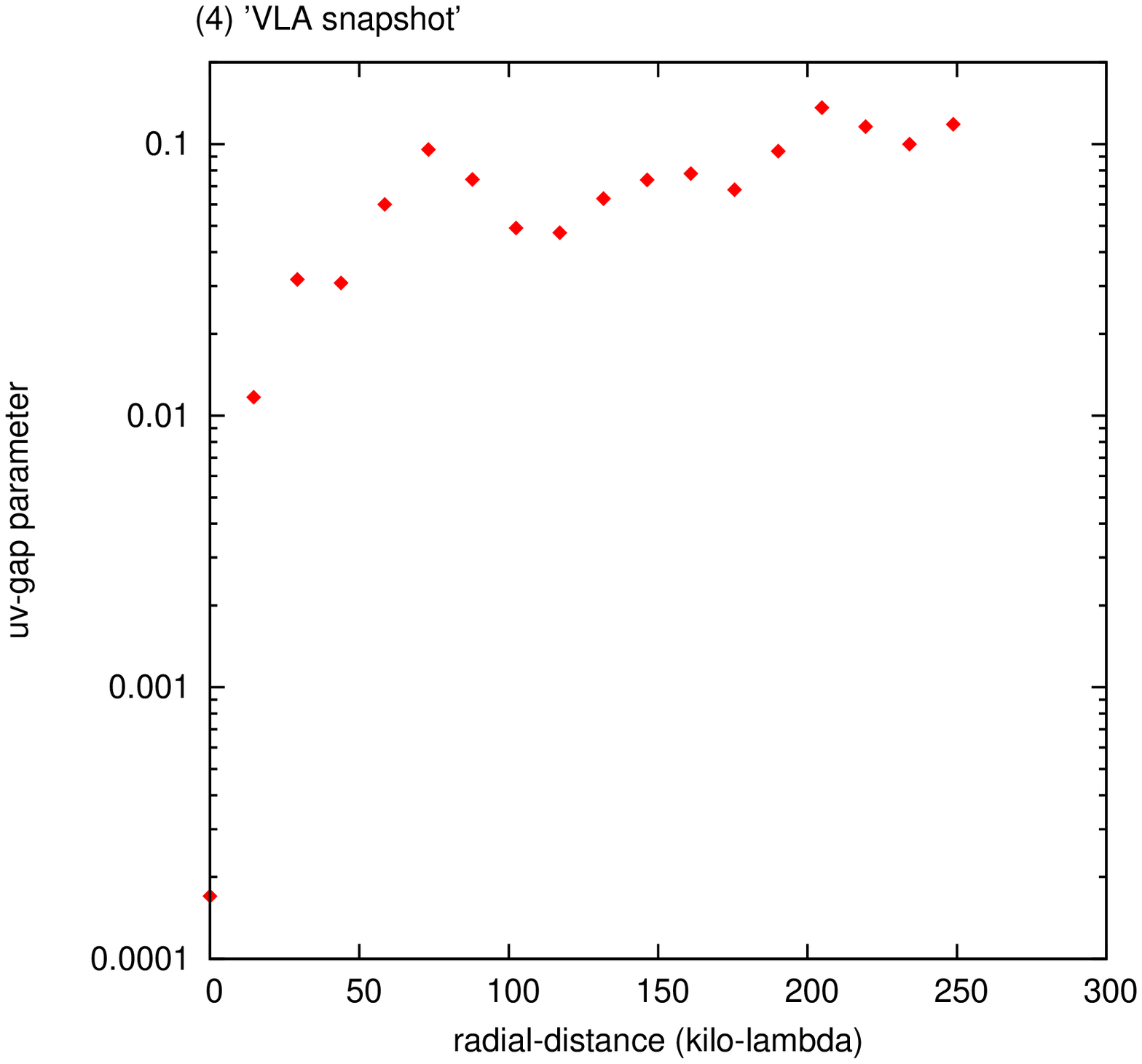} \\
\\
\includegraphics[height=5.58cm]{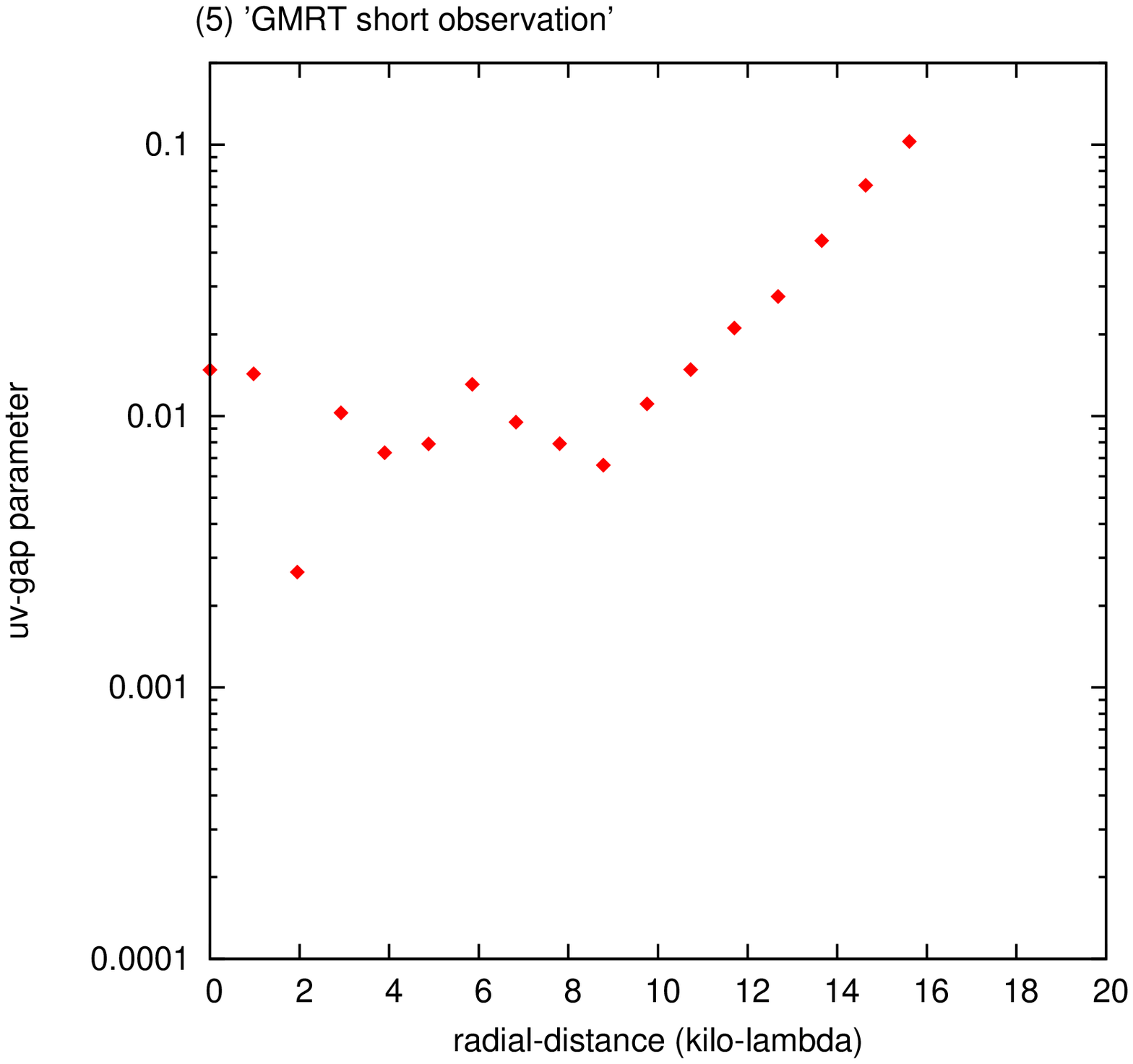} &
\includegraphics[height=5.58cm]{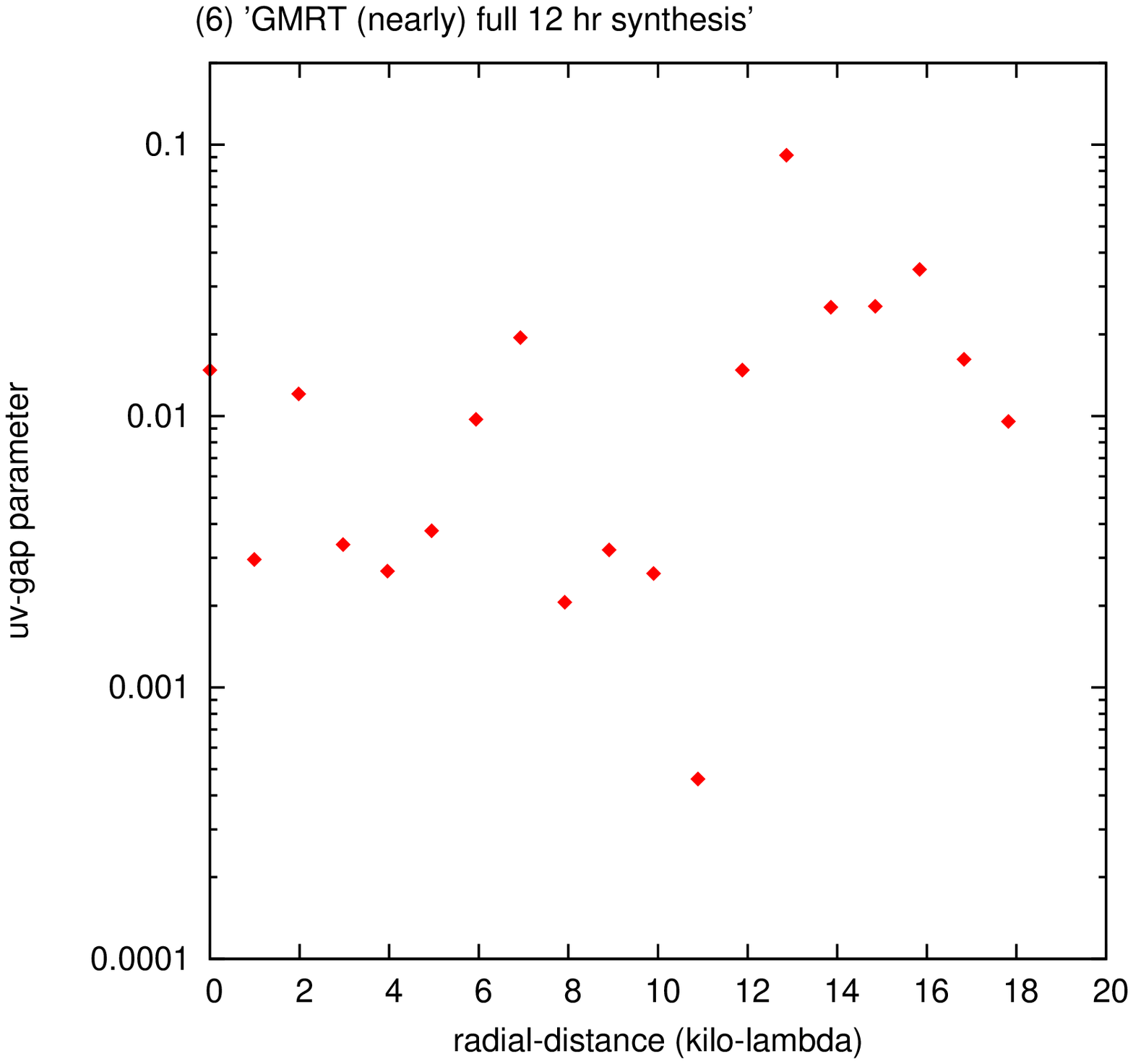}
\end{tabular}
\end{center}
\caption{Azimuthally averaged profiles of $\Delta u/u$ from the density plots of
$\frac{\Delta u}{u} (u, \phi)$ shown in Fig. 7 for all six cases.}
\label{azimuth_profile}
\end{figure}

\newpage

\section*{Appendix 1: An algorithm for calculating $\Delta u/u$ from {\em uv}-coverages}

\subsection*{Initial settings}

\noindent
Let us suppose we have a file containing visibilities or a {\em
uv}-coverage corresponding to an observation with a target array
configuration. The {\em uv}-coverage is given in polar {\em
uv}-coordinates $(u, \theta)$ and spans a range $(u_\mathrm{min},
u_\mathrm{max})$ of {\em uv}-distances. Note that the {\em
uv}-coverage can be completely arbitrary: {\em i.e.}, it can have
asymmetries and gaps along certain position angles --- this does not
affect the calculation. 

The calculation must be done for $N$ individual intervals in position
angle.  The number $N$ is determined by the dynamic range
specification for the array so that $N = {\mathrm SNR}^{1/2}$. This
corresponds to a width of $\Delta \theta = 180/\mathrm{SNR}^{1/2}$
degrees for an individual interval and, respectively for a time
integration of $\Delta \tau = 720/\mathrm{SNR}^{1/2}$ minutes. For a
target $\mathrm{SNR} \rightarrow 10^6$ specified for the SKA, one can
assume (for convenience of calculation)

\medskip
\centerline{
\fbox{
$
N=720\,, \quad\quad\Delta\tau = 1\, \mathrm{min}\,, \quad\quad
\Delta\theta = 0.25\, \mathrm{deg}\,.
$
}
}

\medskip\noindent
The {\em uv}-tracks on individual baselines should either be generated
with an integration time $\Delta \tau$ or averaged to this time (see
Step~2 of the calculation).

\medskip\noindent
For convenience of notation, let us denote 

\medskip
\centerline{
\fbox{
$\delta u \equiv \Delta u/u$
}}

\medskip\noindent
in the
following description of the calculation algorithm.

\subsection*{Calculation of $\Delta u/u$}

\begin{itemize}

\item[Step~~1.] Break the Fourier plane $(u,\theta)$ into $N$
intervals in $\theta$ ($\theta$-intervals) and $N$ intervals in $u$ ($u$-intervals), thus forming an $N\times N$ grid in $(u,\theta)$ with individual grid cells $(u^\prime_i,\theta^\prime_j)$, with $i=1,N$ and $j=1,N$. Note that $\theta \in
(0^\circ,180^\circ)$ and $u \in (0,u_\mathrm{max})$.

\item[Step~~2.] If needed, average the {\em uv}-data to $\Delta \tau$
minutes to ensure having one {\em uv}-point per baseline per
$\theta$-interval.

\item[Step~~3.] In a given  $\theta$-interval, $\theta_j$ ($j=1,N$), add the zero-spacing point
($u_0=0$) to the {\em uv}-points falling into this interval and
calculate (ungridded) values of $\delta u$ from
\[
\delta u_k = (u_k - u_{k-1})/u_k\,,
\]
where $k=1,M$ and $M$ is the total number of {\em uv}-points falling within this $\theta$-interval.
 
\item[Step~~4.] For the given $\theta$-interval, $\theta_j$, map the values $\delta u_k$ to individual grid cells $(u^\prime_i, \theta^\prime_j)$ as follows:
\[
\forall u^\prime_i \in (u_{k-1},u_k): \quad\quad \delta u_i = \delta u_k, \quad\quad k=1,M,\,\,\, i=1,N\,.
\]

\medskip\noindent 
$\bullet$~For each grid cell, check the following two conditions: 

\noindent
$\rightarrow$~If an $i$-th grid cell contains one of the ungridded values
, $\delta u_k$, then set its {\em uv}-gap value to $\delta u_k$
--- that is, no interpolation between the $\delta u_k$ and
$\delta u_{k+1}$ is needed for this grid cell.

\noindent
$\rightarrow$~If an $i$-th grid cell contains several ungridded values of 
$\delta u$, set its $\delta u$ value by  averaging these $\delta u$
values.

\medskip\noindent 
$\bullet$~After mapping the last ($M$-th) ungridded value of $\delta u$, check 
the following condition:

\noindent
$\rightarrow$~If $u_M < u_\mathrm{max}$, assign $\delta u =1$ to the grid
cells falling within the interval $(u_M,u_\mathrm{max})$, starting from the 
next grid cell after the cell containing the {\em uv}-point $u_M$.  

\medskip\noindent
As a result of this Step,
the given $\theta$-interval is divided into $N$ grid cells in $u$ each
assigned a value of $\delta u$.

\item[Step~~5.] Repeat Steps 3 and 4 for all $\theta$-intervals and
populate the entire $(u^\prime,\theta^\prime)$ grid with {\em uv}-gap values
$\delta u_{i,j}$, where $i=1,N$ is the index in $u$-intervals and
$j=1,N$ is the index in $\theta$-intervals.

\item[Step~~6.] Plot the gridded distribution $\delta u_{i,j}$ for
visual inspection, in the rectangular coordinate system
$(u,\theta)$. No smoothing in $u$ or $\theta$ is required.

\item[Step~~7.] Calculate $N$ azimuthal averages of $\delta u$
\[
{\langle\delta u\rangle}_i = \frac{1}{N}\sum_{j=1}^{N} \delta u_{i,j}
\] 
and their variances
\[
\sigma^2_i = \frac{1}{N} \sum_{j=1}^{N} (\delta u_{i,j} - {\langle\delta u\rangle}_i )^2\,.
\]

\item[Step~~8.] Plot the resulting one-dimensional distribution
$(u^\prime_i,{\langle\delta u\rangle}_i)$, with the respective dispersions, $\sigma_i$ as errorbars. This gives a radial
profile of the {\em uv}-gap.

\item[Step~~9.] Calculate the integral FoM value of $\delta u$ and its
variance from
\[
\langle\Delta u/u\rangle_{u,\theta} = \frac{1}{N}\sum_{i=1}^{N} \langle\delta u\rangle_{i}\,,
\]
\[
\sigma^2_{\Delta u/u} =  \frac{1}{N}\sum_{i=1}^{N} \sigma^2_{i}\,.
\]

\end{itemize}

\noindent
Various refinements to the procedure described in above can be
made. One possibility, for instance, is to calculate $\Delta u/u$
within an angular sector with the width of $\Delta \theta$, then shift
this sector by an angle corresponding to the observation intergration
time and repeat this procedure until the entire range of the position
angles is covered. The resulting distribution of $\Delta u/u$ can be
then averaged or smoothed over $\Delta \theta$.

\newpage

\section*{Appendix 2: A set of figures of merit for SKA configuration evaluation}

\noindent
This appendix describes a proposal for an initial set of figures of
merit (FoM) for evaluating imaging performance of different array
configurations for the SKA. The set of FoM proposed includes general
performance metrics and does not include specific requirements on
array configuration coming from individual SKA Key Science Projects.
The critical issues for evaluating imaging
performance provided by a given array configuration are:

\begin{itemize}
\item[1.]~Providing optimal shape of the point-spread-function (PSF). 
\item[2.]~Minimizing sidelobes of the PSF. 
\item[3.]~Evaluating sensitivity to all spatial scales sampled by the SKA.
\end{itemize}

Table~A1 sumarizes a basic set of seven FoM suitable for evaluating the
three conditions listed above. The FoM listed in Table~1 are decribed
in polar coordinates in the {\em uv}-plane $(u,\theta)$ and image
plane $(r,\theta)$.

\begin{table}[h]
\centerline{\small {\bf Table A1:} A set of figures of merit for SKA configuration evaluation}
\begin{tabular}{||l|l|ll|l||}\hline\hline
\multicolumn{1}{||c}{FoM} & \multicolumn{1}{|c}{Description} & \multicolumn{2}{|c}{Name} & \multicolumn{1}{|c||}{Goal} \\\hline\hline
Point Spread & Major Axis & BMA, & $b_\mathrm{maj}$ &                           \\
Function (PSF)  & Minor Axis & BMI, & $b_\mathrm{min}$ &  $b_\mathrm{min}/b_\mathrm{maj}\rightarrow 1$ \\ 
  & P.A. of Major Axis & PAN,  & $\theta_\mathrm{maj}$ &  \\
  & PSF Shape & PSF, & $B(r,\theta)$ & $B(r,\theta) \rightarrow \mathrm{Gaussian}$\\\hline
PSF Sidelobes & Maximum Positive Sidelobe & MPS, & $\sigma_\mathrm{+}$ & $\sigma_\mathrm{+} \rightarrow 0$ \\
 & Maximum Negative Sidelobe & MNS, & $\sigma_\mathrm{-}$ & $\sigma_\mathrm{-} \rightarrow 0$ \\
  & Sidelobe RMS & RMS, & $\sigma_\mathrm{rms}$ & $\sigma_\mathrm{rms} \rightarrow 0$ \\ \hline
UV Gap & Integrated Value   & UVG, & $\Delta u/u$ & $\langle \Delta u/u \rangle_{u,\theta} \rightarrow 0$ \\
  & Dispersion & UVD, & $\Delta u/u(u,\theta)$ & $\sigma_{\Delta u/u} \rightarrow 0$ \\\hline\hline
\end{tabular}
\end{table}


\begin{thebibliography}{}

\bibitem{Borkowski} Borkowski, T.J. (1989) "Accurate algorithm to transform
geocentric to geodetic coordinates", Bull. G\'eod., vol. 63, p. 50

\bibitem{Bregman} Bregman, J.D. (2000) "Concept design for low frequency array",
SPIE proceedings: Astronomical telescopes and
instrumentation, radio telescopes, vol. 4015, p. 1

\bibitem{Bregman1} Bregman, J.D. (2005) "Cost effective frequency ranges for multi-beam
dishes, cylinders, aperture arrays, and hybrids", Experimental astronomy vol. 17,
pp. 407--416

\bibitem{Bridle} Bridle, A.H. \& Schwab, F.R. (1999) "Bandwidth and time-average smearing", ASP Conf. Series, vol. 180, p. 371 

\bibitem{burke} Burke, B. \& Graham-Smith, F. (2002) "An introduction to radio
astronomy", Cambridge (2nd edition)

\bibitem{clark80} Clark, B. (1980) "An efficient implementation of the algorithm
CLEAN", A\&A, vol. 89, p. 377

\bibitem{CHW04} Cohanim, B.E., Hewitt, J.N. \& de~Weck, O. (2004) "The design of
radio telescope array configuration using multiobjective optimization:
Imaging performance versus cable length", ApJSS vol. 15, pp. 705--719

\bibitem{Conway} Conway, J. (1998) "Self-spiral geometries for the LSA/MMA",
ALMA memo 216

\bibitem{Conway2} Conway, J. (2000a) "Observing efficiency of strawperson zoom array",
ALMA memo 283

\bibitem{Conway3} Conway, J. (2000b) "A possible layout for a spiral zoom array
incorporating terrain constraints", ALMA memo 292

\bibitem{cornwell} Cornwell, T.J. (1986) "Crystalline antenna arrays", MM array memo 38

\bibitem{cornwell92} Cornwell, T.J. \& Perley, R.A. (1992)
"Radio-Interferometric Imaging of Very Large Fields", A\&A, vol. 261, p. 353

\bibitem{fukushima} Fukushima, T. (2006) "Transforming from Cartesian
to Geodetic Coordinated Accelerated by Halley's Method", Journal of
Geodesy, vol. 79, pp. 689--693

\bibitem{HM67} Heiskanen W.A. \& Moritz, H. (1967) "Physical geodesy", WH Freeman,
New York

\bibitem{Memo 69} International SKA Project Office (2006)
"Reference design for the SKA", SKA memo 69

\bibitem{memo45} Jones, D.L. (2003) "SKA science requirements",
SKA memo 45.

\bibitem{Kogan} Kogan, L. (2000a) "The imaging characteristics
of an array with minimum side lobes", ASP Conf. Series, vol. 217, p. 348

\bibitem{Kogan} Kogan, L. (2000b) "Optimizing a large array configuration to minimize
the side lobes", IEEE transactions on antennas and propagation, vol. 48, p. 1075

\bibitem{KC05} Kogan, L. \& Cohen, A. (2005) "Optimization of the LWA antenna station
configuration minimising side lobes", LWA memo 21


\bibitem{lobanov2001} Lobanov, A.P., Gurvits, L.I., Frey, S., Schilizzi, R.T.,
 Kawaguchi, N. \& Pauliny-Toth, I.I.K. (2001) "VLBI Space Observatory Programme
Observation of the Quasar PKS 2215$+$020: A New Laboratory for Core-Jet Physics at
z=3.572", ApJL, vol. 547, pp.~714--721

\bibitem{lobanov2003} Lobanov, A.P. (2003) "Imaging with the SKA: Comparison to other
future major instruments", SKA memo 38

\bibitem{Lonsdale1} Lonsdale, C.J., Doelman S.S. \& Oberoi, D. (2004)
"Efficient imaging strategies for next-generation radio arrays", Experimental
Astronomy, vol. 17, pp. 345--362

\bibitem{Lonsdale} Lonsdale, C.J. (2005) "Configuration considerations for low
frequency arrays", ASP Conf. Series, vol. 345, p. 399

\bibitem{MH05} Morita, K.-I. \& Holdaway, M. (2005) "Array configuration design of the
atacama compact array", ALMA memo 538

\bibitem{Noordam} Noordam, J.E. (2001) "Guidelines for the LOFAR array configuration",
Project LOFAR, ASTRON, The Netherlands \\
{\tt (http://www.astron.nl/$\sim$noordam/lofar/config.ps)}

\bibitem{Okabe} Okabe, A., Boots, B., Sugihara, K. \& Chiu, S.N. (2000)
"Spatial tessellations -- Concepts and applications of Voronoi diagrams",
John Wiley (2nd edition)

\bibitem{osullivan09} O'Sullivan, S.P., Stil, J.M., Taylor, A.R., Ricci, R.,
Grant, J.K. \& Shorten, K. 2009, SKADS Memo 31, www.skads-eu.org

\bibitem{perley99} Perley, R.A. (1999) "High dynamic range imaging",
ASP Conf. Series, vol. 120, p. 275

\bibitem{evlamemo63} Perley, R.A. \& Clark, B. (2003)
"Scaling relations for interferometric post-processing", EVLA memo 63

\bibitem{tms86} Thompson, A.R., Moran, J.M. \& Swenson, G.W., Jr. (1986)
"Interferometry and synthesis in radio astronomy", New York, Wiley.

\bibitem{tms01} Thompson, A.R., Moran, J.M. \& Swenson, G.W., Jr. (2001)
"Interferometry and synthesis in radio astronomy", New York, Wiley (2nd edition)

\bibitem{voronkov} Voronkov, M.A. \& Wieringa, M.H. (2004) "Dynamic range of
the SKA images", SimWG report, Sydney
{\tt (http://www.narrabri.atnf.csiro.au/$\sim$vor010/ska/SimWG\_Report2.pdf)}

\bibitem{memo68} Voronkov, M.A. \& Wieringa, M.H. (2005) "Dynamic range loss due to
the retarded baseline effect", SKA memo 68.

%

\bibitem{memo16} Wright, M.C.H. (2002) "A model for the SKA", SKA memo 16

\bibitem{SKA memo 46} Wright, M. (2004) "SKA Imaging",
         SKA memo 46

\end{thebibliography}
\end{document}